\pdfoutput=1
\documentclass[12pt,a4paper]{article}
\usepackage{jheppub}
\usepackage[usenames,dvipsnames]{xcolor}
 
\usepackage{microtype}
\usepackage{amssymb} 
\usepackage{amsmath}
\usepackage{mathtools}
\usepackage{amsfonts}    
\usepackage{dsfont}
\usepackage{pdfpages}
\usepackage{verbatim}
\usepackage{braket}
\usepackage{bm}
\usepackage{tensor}
\usepackage{upgreek}
\usepackage{subcaption}
\usepackage{simpler-wick}
\usepackage{enumitem}
\usepackage[hang,flushmargin]{footmisc}
\usepackage{amsthm}
\usepackage{tcolorbox}
\usepackage{nicematrix}
\usepackage[table]{xcolor}
\usepackage{fontawesome}
\usepackage{framed}

\usepackage{tikz}
\usepackage{listings}
\usepackage{color}
\usetikzlibrary{decorations.markings}

\usepackage{pgfplots}
\usepgfplotslibrary{fillbetween}
\pgfplotsset{compat=1.18}

\usepackage{tocloft}
\definecolor{rosewood}{rgb}{0.4, 0.0, 0.04}
\definecolor{pyblue}{RGB}{31, 119, 180}
\definecolor{pyorange}{RGB}{255, 127, 14}
\definecolor{pygreen}{RGB}{44, 160, 44}
\definecolor{pyred}{RGB}{214, 39, 40}
\definecolor{lightgray}{gray}{0.9}
\colorlet{lightpyblue}{pyblue!30!white}
\colorlet{lightpyred}{pyred!30!white}
\colorlet{lightpygreen}{pygreen!30!white}

\usepackage{hyperref}
\hypersetup{
    pdfencoding=unicode,
    colorlinks=true,
    urlcolor=rosewood,
    linkcolor=pyblue,
    citecolor=pygreen,
    pdftitle={This is the title},
    pdfauthor={Denis Werth},
    pdfdisplaydoctitle=true,
    pdfstartview=FitH,
    linktocpage=true
}

\def \d {\mathrm{d}}

\def \p {\bm{p}}

\def \x {\bm{x}}
\def \s {\bm{s}}
\def \z {\bm{z}}

\def \G {\mathcal{G}}

\def \I {\mathcal{I}}

\def \K {\mathcal{K}}

\def \M {\mathcal{M}}

\def \O {\mathcal{O}}

\def \S {\mathcal{S}}

\def \SO {\mathrm{SO}}
\def \Li {\mathrm{Li}}
\def \AdS {\mathrm{AdS}}

\def \Im {\mathrm{Im}}
\def \Re {\mathrm{Re}}

\def \Res {\mathrm{Res}}

\newcommand{\lla}{\langle \! \langle}
\newcommand{\rra}{\rangle \! \rangle}

\title{Analytic Continuation of Conformal Integrals in Momentum Space}

\author[1, 2]{Jonathan Gr\"afe,}\emailAdd{graefe@mpp.mpg.de}
\author[1]{Prashanth Raman}\emailAdd{praman@mpp.mpg.de}
\author[1, 2]{and Denis Werth}\emailAdd{werth@mpp.mpg.de}

\affiliation[1]{Max Planck Institute for Physics, Werner-Heisenberg-Institut, Munich, D-85748, Germany}
\affiliation[2]{Max Planck-IAS-NTU Center for Particle Physics, Cosmology and Geometry}

\abstract{Conformal symmetry strongly constrains correlation functions. In momentum space, the conformal Ward identities are solved for three-point functions by integrals of a product of three Bessel functions (``triple-$K$ integrals''); more generally, integrals of this type (``multiple-$K$ integrals''), serve as the building blocks of a wide class of higher-point conformal correlators. These integrals belong to the class of generalised hypergeometric functions, and while their series representations are known in principle, none converges throughout the entire physical region of kinematic space selected by momentum conservation. In this work, we construct series representations adapted to exactly this physical domain. Using the method of brackets, which turns the evaluation of definite integrals into solving a linear system of algebraic equations, we derive various Lauricella-type series representations for multiple-$K$ integrals. For triple-$K$ integrals, conventionally expressed in terms of the Appell $F_4$ function whose series converges only outside the triangle-inequality region, we find instead a compact, two-branch series that converges throughout the entire physical region and for arbitrary scaling dimensions. We then extend this construction to general multiple-$K$ integrals: by introducing a new set of kinematic variables, we build an iterative series representation that converges for all physical kinematic configurations.
}

\begin{document}

\setcounter{tocdepth}{3}
\maketitle
\setcounter{page}{1}

%-------------------------------------
%-------------------------------------
%-------------------------------------
\newpage
\section{Introduction}

Conformal symmetry imposes powerful constraints on the form of correlation functions: two- and three-point functions are fixed entirely up to a handful of constants, and the general form of position-space $n$-point conformal correlators has been understood long ago~\cite{Polyakov:1970xd, Schreier:1971um, Osborn:1993cr}. Yet the form of the answer depends crucially on the space in which one chooses to work. In position space, a rich set of techniques has been developed to compute and organise conformal correlators, many of them mirroring tools familiar from the study of Feynman integrals. Star-triangle relations turn certain conformal integrals into pure algebra and underlie the theory's integrability~\cite{Symanzik:1972wj, Kazakov:1983dyk, Chicherin:2012yn}, while Symanzik's star formula exchanges position-space cross-ratios for conjugate Mellin variables in which conformal symmetry becomes manifest~\cite{Mack:2009mi, Penedones:2010ue}. Embedding space, meanwhile, linearises conformal covariance much as spinor-helicity variables do for scattering amplitudes~\cite{Dirac:1936fq, Weinberg:2010fx, Costa:2011mg}. Eventually, contact Witten diagrams evaluate to $D$-functions, whose dimension-shift identities in the operator scaling dimensions closely parallel integration-by-part relations for Feynman integrals~\cite{Witten:1998qj, DHoker:1999mqo}.

\paragraph{Momentum-space conformal integrals.} In momentum space, conformal generators are no longer algebraic but act as genuine (partial) differential operators on correlators. Solving these conformal Ward identities is a complicated task, as solutions generically take the form of generalised hypergeometric functions of momentum ratios. An important class of solutions to these Ward identities can be conveniently represented by integrals over a product of Bessel-$K$ functions---so-called ``multiple-$K$ integrals''---, of the form~\cite{Bzowski:2013sza, Coriano:2013jba, Bzowski:2015pba, Bzowski:2015yxv, Coriano:2019sth, Coriano:2020ccb, Coriano:2020ees}:
\begin{equation*}
    \int_0^\infty \d z z^{\alpha-1} \prod_{j=1}^n \textcolor{pyblue}{p_j}^{\Delta_j-\frac{d}{2}} K_{\Delta_j-\frac{d}{2}}(\textcolor{pyblue}{p_j} z) \,, \qquad 
    \vcenter{\hbox{
    \begin{tikzpicture}[line width=1. pt, scale=1.5]

        \foreach \i in {0,...,5}
        \coordinate (V\i) at ({60*\i}:0.8);
        
        \fill[gray!10] (V0) -- (V1) -- (V2) -- (V3) -- (V4) -- (V5) -- cycle;
        
        \draw[black] (V0) -- node[midway, above right, yshift=-3pt, xshift=3pt] {\textcolor{pyblue}{$p_n$}} (V1);
        \node[black] at ($(V1)!1/2!(V2)$) {$\cdots$};
        \draw[black] (V2) -- node[midway, above left, yshift=-3pt, xshift=-3pt] {\textcolor{pyblue}{$p_4$}} (V3);
        \draw[black] (V3) -- node[midway, below left, yshift=3pt, xshift=-3pt] {\textcolor{pyblue}{$p_3$}} (V4);
        \draw[black] (V4) -- node[midway, below, yshift=-3pt] {\textcolor{pyblue}{$p_2$}} (V5);
        \draw[black] (V5) -- node[midway, below right, yshift=3pt, xshift=3pt] {\textcolor{pyblue}{$p_1$}} (V0);

    \end{tikzpicture} 
    }} \,,
\end{equation*}
where $\Delta_j$ are the operator scaling dimensions ($j=1, \ldots, n$), and $p_j\equiv |\p_j|$ are the magnitudes of the $d$-dimensional (Euclidean) external momenta, which close by momentum conservation. Since the holographic bulk-to-boundary propagator is itself built from a Bessel-$K$ function, these integrals are nothing but Fourier transforms of position-space $D$-functions for contact Witten diagrams~\cite{Raju:2012zr, Albayrak:2018tam, Albayrak:2020isk, Bzowski:2022rlz}, and can equivalently be recast as Feynman integrals with integrated momenta running along the edges of a simplex~\cite{Bzowski:2019kwd, Bzowski:2020kfw}. For $n=3$, these ``triple-$K$'' integrals govern all three-point scalar, spinning and tensorial conformal correlators~\cite{Bzowski:2013sza, Isono:2019ihz, Isono:2019wex, Jain:2021wyn, Gillioz:2020wgw, Bautista:2019qxj}. In the context of cosmology, where conformal symmetry emerges on the future boundary of de Sitter spacetime, multiple-$K$ integrals are the fundamental building blocks for cosmological correlators~\cite{Belrhali:2026ktb, Belrhali:2026rkn, Belrhali:2026act, Grafe:2026qsm, Grafe:2026avi}. And in the context of Feynman integrals, they have in fact been present all along without being recognised as such: the one-loop massless triangle diagram is itself a triple-$K$ integral in disguise~\cite{Boos:1990rg, Davydychev:1992xr}, with closely related integrals also appearing in configuration-space banana diagrams~\cite{Groote:2018rpb, Groote:2005ay, Cacciatori:2023tzp}.

\vskip 4pt
Multiple-$K$ integrals are known to evaluate to generalised multivariable hypergeometric functions. However, the corresponding series representations typically have a limited radius of convergence and, in particular, fail to converge throughout the physical region of kinematic space selected by momentum conservation. A paradigmatic example is provided by the triple-$K$ integral which describes all three-point conformal correlators:
\begin{equation*}
    \lla \O_{\Delta_1}(\p_1) \O_{\Delta_2}(\p_2) \O_{\Delta_3}(\p_3) \rra = C_{123} \, \int_0^\infty \d z z^{\alpha-1} \prod_{j=1}^3 p_j^{\Delta_j-\frac{d}{2}} K_{\Delta_j-\frac{d}{2}}(p_j z) \,,
\end{equation*}
where the double bracket notation means that we have stripped off the overall momentum-conserving $\delta$-function. For arbitrary scaling dimensions, the triple-$K$ integral can be expressed in terms of the Appell $F_4$ hypergeometric function of the two dimensionless kinematic ratios $(p_1/p_3)^2$ and $(p_2/p_3)^2$. The associated series representations, however, only converge outside of the triangle inequalities: for $p_1+p_2<p_3$. Even  the analytic continuations obtained by expanding around the singular loci of the corresponding differential equations do not provide a convergent representation throughout the physical domain. Faced with this, the only recourse so far has been numerical integration~\cite{Bzowski:2020lip, Banik:2026smq}, although some analytic continuations have been derived in~\cite{Alkofer:2008dt, Ananthanarayan:2019icl, Ananthanarayan:2020xut}. More generally, momentum-space representations of conformal correlators appear to sit in tension with physical kinematics: the two seem almost mutually exclusive, and analytically continuing momentum-space conformal integrals into the physical region has remained a long-standing open problem.

\vskip 4pt
In this paper, we derive a new class of series representations for multiple-$K$ integrals that converge throughout the entire physical kinematic region carved out by momentum conservation. By introducing a suitable set of dimensionless kinematic variables and constructing solutions to the conformal Ward identities iteratively, we obtain series representations that converge absolutely for all physical kinematic configurations. We also systematically explore alternative representations of multiple-$K$ integrals using the method of brackets, which naturally singles out dimensionless variables adapted to their underlying integral structure. In particular, we derive type-$C$ and type-$A$ Lauricella series representations in terms of the ratios $u_i \equiv p_i/p_n$ ($i=1, \ldots, n-1$) and $x_j\equiv 2p_j/S_n$ ($j=1, \ldots, n$) with $S_n\equiv p_1+\cdots+p_n$ the total energy. These results open the door to a unified analytic treatment of momentum-space conformal correlators directly in their physical domain.

\paragraph{Method of brackets.} Analytic continuation of special functions defined as solutions of differential equations is naturally performed by constructing local expressions around singular loci. For example, the Gauss hypergeometric function ${}_2F_1$ admits standard series representations around $z=0$, $z=1$ and $z=\infty$. Obtaining an expression around a generic point, however, typically requires numerically integrating the differential equation along a path connecting the neighbourhood of a singular locus to the point of interest. The {\it method of brackets}, which is based on Ramanujan's master theorem, provides a powerful alternative: it allows one to derive series representations of definite integrals directly around generic points~\cite{HALLIDAY1987241, Dunne:1987am, Dunne:1987qb, gonzalez2008definiteintegralsmethodbrackets, gonzalez2010methodbrackets2examples, Gonzalez:2021vqh, Koutschan2011TheII}, see~\cite{GONZALEZ2015214, Gonzalez2017, Ananthanarayan:2021not, Ghosh:2022net} for applications to Feynman integrals. At its core, the method reduces the problem of finding series representations to definite integrals to solving a linear system of algebraic equations. Schematically, one first expands the integrand around the origin and associates to the integral a formal bracket series. For example,
\begin{equation*}
    \int_0^\infty \d z z^{\alpha-1} e^{-uz} K_\nu(z) \,\, \widehat{=} \,\, \sum_\pm \sum_{m, n=0}^\infty \, c_{m, n}^\pm(\nu) \, \textcolor{pyblue}{\langle m+2n\pm\nu+\alpha \rangle}\,, \quad (u>0) \,,
\end{equation*}
where $c_{m, n}^\pm(\nu)$ are the coefficients arising from the formal expansion of the integrand around $z=0$. The bracket $\langle m+2n\pm\nu+\alpha \rangle$ is a compact object {\it representing} the divergent integral $\int_0^\infty \d z z^{m+2n\pm\nu+\alpha-1}$ (this explains the use of the notation ``$\widehat{=}$'' instead of ``$=$''). A well-defined representation of the original integral is then obtained by evaluating the bracket series ``on shell'', namely on the solution of the bracket equation: $m+2n\pm\nu+\alpha=0$. Since this system has rank one, one may solve for either $m$ or $n$, leading to two distinct series representations, which correspond to different analytic continuations of the same function. We apply this framework systematically to multiple-$K$ integrals, starting from different integral representations of the Bessel-$K$ function, and find new Lauricella-type series representations. 

\paragraph{Analytic continuation to the physical region.} The fundamental obstacle to constructing convergent series representations of multiple-$K$ integrals throughout the physical region is that the physical kinematic domain is unbounded when expressed in the standard momentum ratios $u_i\equiv p_i/p_n$ ($i=1, \ldots, n-1$), and does not contain the origin. The key insight underlying our construction is therefore to perform a judicious change of variables that maps the physical kinematic domain onto a bounded region. This allows us to construct series expansions in alternative momentum variables whose domain of convergence encompass the entire physical region. As an illustration, for the triple-$K$ integral, our change of variables maps the unbounded wedge in the standard momentum ratios to the bounded unit square:
\begin{center}
    \begin{tikzpicture}[line width=1. pt, scale=2]

        %Axis
        \draw[->] (-0.1,0) -- (1.5,0) node[right] {$p_1/p_3$};
        \draw[->] (0,-0.1) -- (0,1.5) node[above] {$p_2/p_3$};

        \draw[->] (3.9,0) -- (5.5,0) node[right] {$p_1/S_2$};
        \draw[->] (4,-0.1) -- (4,1.5) node[above] {$p_3/S_2$};

        % Ticks and labels on x-axis
        \foreach \x in {1} {
            \draw (\x, 0.03) -- (\x, -0.03) node[below] {\small $\x$};
        }
        \foreach \y in {1} {
            \draw (0.03, \y) -- (-0.03, \y) node[left] {\small $\y$};
        }
        \node[below left] at (0,0) {\small $0$};

        \foreach \x in {5} {
            \draw (\x, 0.03) -- (\x, -0.03) node[below] {\small $1$};
        }
        \foreach \y in {1} {
            \draw (4.03, \y) -- (3.97, \y) node[left] {\small $1$};
        }
        \node[below left] at (4,0) {\small $0$};

        %Physical region
        \fill[black!5] (0,1) -- (0.5,1.5) -- (1.5,1.5) -- (1.5, 0.5) -- (1, 0) -- cycle;
        \draw[pyblue, line width=1.2pt] (0.5,1.5) -- (0,1) -- (1, 0) -- (1.5, 0.5);

        \fill[black!5] (4,0) -- (4,1) -- (5,1) -- (5, 0) -- cycle;
        \draw[pyblue, line width=1.2pt] (4,0) -- (4,1) -- (5,1) -- (5,0) -- cycle;
        
        \draw[<->, thick, black, line width=1.5pt] (2,0.75) to[bend left=20] (3.4,0.75);
        
    \end{tikzpicture}
\end{center}
with $S_2\equiv p_1+p_2$. The \textcolor{pyblue}{blue} lines indicate the singular loci of the conformal Ward identities. For multiple-$K$ integrals, we implement this idea recursively: at each step, we project the Bessel-$K$ function onto its two independent Bessel-$I$ branches, and reduce the resulting expressions to a terminal two-$K$ integral, which we evaluate using the method of brackets. This procedure naturally generates convergent series representations in the new bounded kinematic variables. Our solutions satisfy the conformal Ward identities by construction, and reduce to the known conformally coupled limit after non-trivial manipulations. 

\paragraph{Outline.} The organisation of this paper is as follows: In Sec.~\ref{sec: conformal integrals}, we review properties of conformal correlators and define conformal integrals, which are the objects of interest for this paper. In Sec.~\ref{sec: method of brackets}, we introduce the method of brackets as the main tool for our goal of obtaining alternative series representations for conformal integrals. We provide some applications of this method in~\ref{subsec: preliminary applications}. We use the method of brackets in Sec.~\ref{sec: different representations} to derive various representations for multiple-$K$ integrals based on different integral representations of Bessel-$K$ functions. Finally, in Sec.~\ref{sec: analytic continuation} we obtain series representations that converge in the entire physical region of the kinematic plane---providing analytic continuations of the known series representations for physical applications. We check that our results still fulfil the conformal Ward identities in~\ref{subsubsec: conformal symmetry} and discuss the conformally coupled limit in~\ref{subsubsec: conformally coupled limit}. Eventually, we provide more details for the triple-$K$ integral in~\ref{subsubsec: triple-K}. Our conclusions are summarised in Sec.~\ref{sec: conclusion}.

\subsection*{Summary of results}
\addcontentsline{toc}{subsection}{Summary of results}

For the reader's convenience, we provide below a summary of our main results, highlighted by boxed equations.

\begin{itemize}
    \item We study the general class of multiple-$K$ integrals defined in~\eqref{eq: def conformal multi-K integral}, with a detailed case study of the triple-$K$ integral~\eqref{eq: def triple-K integral}, which describes all conformal three-point functions of scalar primaries in momentum space.
    \item Eq.~\eqref{eq: triple-K F4 series rep} is the standard Appell $F_4$ representation for the triple-$K$ integral. When one leg is conformally coupled, this reduces to the closed form given in Eq.~\eqref{eq: seed convergent series} which converges in the physical regime after resummation.
    \item Eq.~\eqref{eq: Lauricella F_C rep} is the generalised Lauricella $F_C^{(n-1)}$ representation of multiple-$K$ integrals. These integrals also admit a Lauricella $F_A^{(n)}$ representation, Eq.~\eqref{eq: F_A series rep}.
    \item We derive a new series representation for multiple-$K$ integrals, Eq.~\eqref{eq: multiple-K convergent series} (Eq.~\eqref{eq: triple-K convergent series} for the triple-$K$ integral), that converges throughout the physical region.
\end{itemize}

The series representations for multiple-$K$ conformal integrals, together with numerical checks are collected in a supplemental \textsf{Mathematica} notebook, available at the GitHub repository~[\href{https://github.com/deniswerth/Conformal-Integrals}{\faGithub}].

\paragraph{Notation \& convention.} Spatial $d$-dimensional vectors are written in boldface, $\p$. Our Fourier convention is
\begin{equation}
    f_{\p} = \int \d^d\x \, f(\x) \, e^{i \p\cdot \x} \,, \quad f(\x) = \int \frac{\d^d \p}{(2\pi)^d} \, f_{\p} \, e^{-i \p\cdot\x} \,.
\end{equation}
To avoid cluttered notations throughout the main text, we introduce the following shorthand notation for the product of $\Gamma$-functions:
\begin{equation}
    \Gamma[a_1, \ldots, a_n] \equiv \Gamma(a_1) \cdots \Gamma(a_n) \,, \quad \Gamma[a\pm b] \equiv \Gamma(a+b)\Gamma(a-b) \,.
\end{equation}
To make notations compact, we also use multi-index vectors,~e.g., $\bm{\nu} \equiv (\nu_1, \ldots, \nu_n)$ with the scalar shorthand $|\bm{\nu}| \equiv \sum_{j=1}^n\nu_j$ ($n>0$). For component-wise products, it should be understood that,~e.g.
\begin{equation}
    \p^{\bm{\nu}} \equiv \prod_{j=1}^n p_j^{\nu_j} \,, \quad \frac{(-1)^{\bm{m}}}{\bm{m}!} \equiv \prod_{j=1}^n \frac{(-1)^{m_j}}{m_j!}\,, \quad \Gamma[\bm{\nu}] \equiv \Gamma[\nu_1, \ldots, \nu_n] = \prod_{j=1}^n \Gamma(\nu_j) \,.
\end{equation}
Details about special functions used in this work can be found in App.~A of~\cite{Grafe:2026avi}.

%-------------------------------------
%-------------------------------------
%-------------------------------------
\newpage
\section{Conformal Integrals}
\label{sec: conformal integrals}

In this section, we start by reviewing basic properties of conformal correlation functions, both in position and momentum space, and introduce the class of momentum-space conformal integrals that will be the focus of this work.

\subsection{Position space}

Conformal field theories are invariant under the Euclidean conformal group, which is isomorphic to $\SO(d+1, 1)$ in $d$ spatial dimensions (here, we consider $d>2$). This symmetry imposes powerful constraints on correlation functions~\cite{Polyakov:1970xd, Schreier:1971um, Osborn:1993cr, DiFrancesco:1997nk}. 
In what follows, we will consider scalar operators $\O_i(\x_i)$ of scaling dimensions $\Delta_i \equiv \tfrac{d}{2} + \nu_i$. Although unitarity requires the scaling dimensions to be real (and positive), so that $\nu_i \in \mathbb{R}$, we allow the parameters $\nu_i$ to take arbitrary complex values.

\vskip 4pt
The one-point function vanishes identically, while the two-point function is completely fixed by conformal invariance. In position space, it takes the form: 
\begin{equation}
\label{eq: position-space 2pt function}
    \langle\O_1(\x_1)\O_2(\x_2)\rangle =
    \left\{
    \begin{alignedat}{3} 
        &c_\O \, x_{12}^{-2\Delta}\,, & \quad \Delta_1=\Delta_2=\Delta, \\
        &0\,, & \quad \Delta_1\neq\Delta_2 \,,
    \end{alignedat}
    \right.
\end{equation}
where $x_{ij}\equiv |\x_i-\x_j|$. The overall normalisation $c_\O$ can be absorbed into the definition of the operator. The condition $\Delta_1=\Delta_2$ follows non-trivially from the Ward identity associated with special conformal transformations. Similarly, the three-point function is fixed up to a single constant:
\begin{equation}
    \langle\O_1(\x_1)\O_2(\x_2)\O_3(\x_3)\rangle = \frac{c_{123}}{x_{12}^{\Delta_1+\Delta_2-\Delta_3} x_{13}^{\Delta_1-\Delta_2+\Delta_3} x_{23}^{-\Delta_1+\Delta_2+\Delta_3}} \,,
\end{equation}
where $c_{123}$ is the structure constant, which contains dynamical information and cannot be removed by a rescaling of the operators. For $n\geq4$, conformal symmetry no longer fixes correlators uniquely. Since there are $n(n-1)/2$ independent separations $x_{ij}$ but only $n$ independent scaling constraints, the general solution depends on $n(n-3)/2$ independent conformal invariants. In particular, the four-point function takes the form
\begin{equation}
    \langle \O_1(\x_1)\O_2(\x_2)\O_3(\x_3)\O_4(\x_4) \rangle = f(u, v) \prod_{1 \leq i<j \leq 4} x_{ij}^{2\alpha_{ij}} \,,
\end{equation}
where the exponents $\alpha_{ij}$ are related to the scaling dimensions by the relations
\begin{equation}
    \Delta_i = -\sum_{j=1}^4 \alpha_{ij}\,, \quad (i=1, \ldots, 4)\,.
\end{equation}
Without loss of generality, we can set $\alpha_{ij}=\alpha_{ji}$ and $\alpha_{ii}=0$. These conditions do not uniquely determine the $\alpha_{ij}$, reflecting the freedom to redistribute powers among the pairwise distances. The function $f(u, v)$ depends only on the two conformal cross ratios:
\begin{equation}
    u \equiv \frac{x_{12}^2 x_{34}^2}{x_{13}^2 x_{24}^2} \,, \quad v \equiv \frac{x_{13}^2 x_{24}^2}{x_{14}^2 x_{23}^2} \,.
\end{equation}
The generalisation to higher-point correlation functions is straightforward: the correlator is fixed up to an arbitrary function of the corresponding set of conformal cross-ratios.

\subsection{Momentum space}

Similarly, in momentum space, conformal symmetry imposes stringent constraints on correlators. Their general form can be either inferred from solving the corresponding set of Ward identities, or performing the Fourier transform of position-space correlators. As usual, translational invariance implies momentum conservation:
\begin{equation}
    \langle \O_1(\p_1) \cdots \O_n(\p_n) \rangle = (2\pi)^d \delta^{(d)}(\p_1 + \cdots + \p_n) \lla \O_1(\p_1) \cdots \O_n(\p_n) \rra \,,
\end{equation}
where the double-bracket notation indicates that we have stripped off the momentum conserving delta function. The one-point function vanishes, and the two-point function can be determined explicitly from~\eqref{eq: position-space 2pt function} by performing the Fourier transform:
\begin{equation}
    \lla \O(\p) \O(-\p) \rra = c_\O \, \frac{\Gamma(\tfrac{d-2\Delta}{2})}{\pi^{d/2}4^{\Delta}\Gamma(\Delta)} \, p^{2\Delta-d} \,.
\end{equation}
Notice that for $\Delta=\tfrac{d}{2}+k$ with $k\in \mathbb{N}_{\geq0}$, the Fourier transform is ill-defined due to short distance singularities, and requires renormalisation, see e.g.~\cite{Bzowski:2015yxv, Bzowski:2015pba} for a detailed discussion. 

\subsubsection{Conformal Ward identities}

While conformal transformations act naturally in position space, they lead to differential operators in momentum space. Dilatation, being a linear transformation of the position, $\x \to \x'=\x+\delta\x$ with $\delta\x=\lambda\x$, leads to the following first-order differential equation Ward identity:
\begin{equation}
    \left[ \Delta_t - (n-1)d + \sum_{i=1}^n \p_i \cdot \bm{\partial}_i \right] \lla \O_1(\p_1) \cdots \O_n(\p_n) \rra = 0 \,, \quad \text{with} \quad \bm{\partial}_i \equiv \frac{\partial}{\partial \p_i} \,,
\end{equation}
where $\Delta_t=\sum_{i=1}^n\Delta_i$. This constraint implies that the (reduced) correlators are homogeneous functions of degree $\Delta_t-(n-1)d$. Special conformal transformations, however, act non-linearly in position space, with $\delta\x = \bm{b} x^2 - 2\x (\bm{b}\cdot \x)$, and lead to Ward identities which are second-order partial differential equations. The generator for special conformal transformations acting on a leg $\p_i$ is
\begin{equation}
    \bm{\K}_i \equiv \p_i \partial_i^2 - 2(\p_i \cdot \bm{\partial}_i) \bm{\partial}_i + 2(\Delta_i-d)\bm{\partial}_i \,,
\end{equation}
and the $n$-point correlator satisfies 
\begin{equation}
\label{eq: conformal Ward identity}
    \sum_{i=1}^n \bm{\K}_i \, \lla \O_1(\p_1) \cdots \O_n(\p_n) \rra = 0 \,.
\end{equation}
Solving these equations is, in general, a complex task and typically gives rise to generalised hypergeometric functions whose analytic structure remains only partially understood, see e.g.~\cite{Bzowski:2013sza, Coriano:2013jba, Coriano:2019sth, Coriano:2020ccb} for three- and four-point functions, and~\cite{Arkani-Hamed:2015bza, Arkani-Hamed:2018kmz} in the context of cosmology where correlators lying on the space-like future boundary of de Sitter spacetime are constrained by conformal symmetry.

\subsubsection{Three-point functions}

Poincar\'e invariance implies that momentum-space three-point functions depend only on the three momentum magnitudes, $p_j \equiv |\p_j|$ for $j=1, 2, 3$. Further imposing the appropriate scaling behaviour leaves only two independent dimensionless ratios as non-trivial kinematic variables. The remaining constraints from dilatation and special conformal Ward identities form a system equivalent to the differential equations defining a generalised hypergeometric function of two variables (see Sec.~\ref{subsec: triple-K integrals: recovering Appell F4} for details). The general solution can be represented by a triple-$K$ integral:
\begin{equation}
    \lla \O_1(\p_1) \O_2(\p_2) \O_3(\p_3) \rra = C_{123} \, \p^{\bm{\nu}} \, \I^{(3)}_{\alpha, \{\nu_1 \nu_2 \nu_3\}} (p_1, p_2, p_3) \,,
\end{equation}
where we define $\p^{\bm{\nu}} \equiv \prod_{j=1}^3 p_j^{\nu_j}$ using multi-index notation, $C_{123}$ is an integration constant (proportional to $c_{123}$), and
\begin{equation}
\label{eq: def triple-K integral}
    \boxed{
    \I^{(3)}_{\alpha, \{\nu_1 \nu_2 \nu_3\}} (p_1, p_2, p_3) \equiv \int_0^\infty \d z z^{\alpha-1} \prod_{j=1}^3 K_{\nu_j}(p_j z) \,,
    }
\end{equation}
is the triple-$K$ integral~\cite{Bzowski:2013sza}, with $K_\nu$ denoting the modified Bessel-$K$ function. This representation can be obtained by solving the conformal Ward identities through separation of variables and selecting the branch of solutions corresponding to the Bessel-$K$ functions. This choice ensures the absence of unphysical singularities at collinear momentum configurations, e.g.~$p_1+p_2=p_3$~\cite{Coriano:2013jba, Bzowski:2015pba}. A sufficient generic condition for absolute convergence of the master integral~\eqref{eq: def triple-K integral} in the IR ($z=0$) is
\begin{equation}
\label{eq: IR convergence}
    \alpha > |\Re\,\nu_1| + |\Re\,\nu_2| + |\Re\,\nu_3| \,,
\end{equation}
with $p_j>0$ ($j=1, 2, 3$). The exponential decay of the Bessel-$K$ function guarantees convergence in the UV ($z\to\infty$). In some cases, the triple-$K$ integral can be extended beyond the region~\eqref{eq: IR convergence} by analytic continuation~\cite{Bzowski:2015pba, Bzowski:2015yxv, Bzowski:2017poo}.

\paragraph{Example: one-loop triangle integral.} For half-integer $\nu_j$ parameters, the integral~\eqref{eq: def triple-K integral} simplifies since the Bessel-$K$ function reduces to elementary functions. As an example, the master integral $\I^{(3)}_{2, \{000\}}$ is equivalent to the one-loop triangle integral~\cite{tHooft:1978jhc, Davydychev:1992xr}:
\begin{equation}
    \I^{(3)}_{2, \{000\}}(p_1, p_2, p_3) = \frac{1}{4\pi^2} \int \frac{\d^4 \ell}{\ell^2 (\ell-p_1)^2 (\ell+p_2)^2} \,,
\end{equation}
and evaluates explicitly to the Bloch-Wigner function~\cite{Zagier:2007knq}:
\begin{equation}
    \I^{(3)}_{2, \{000\}} = \frac{1}{2\sqrt{-J^2}} \left[\Li_2\,z_+ - \Li_2\,z_- + \frac{1}{2}\log (z_+z_-) \log\left(\frac{1-z_+}{1-z_-}\right)\right] \,,
\end{equation}
where $\sqrt{-J^2} = p_3^2(z_+-z_-)$ is proportional to the area of the triangle with perimeter $p_1+p_2+p_3$, and 
\begin{equation}
    z_\pm \equiv \frac{1}{2}\left(1+u-v \pm \sqrt{(1+u-v)^2 - 4u}\right) \,,
\end{equation}
when choosing the branch $\Im\,z_+\geq0$, with $u\equiv (p_1/p_3)^2=z_+z_-$ and $v\equiv (p_2/p_3)^2=(1-z_+)(1-z_-)$. Other special cases can be found in~\cite{Bzowski:2015pba}. More generally, the triple-$K$ integral~\eqref{eq: def triple-K integral} is equivalent to the one-loop triangle integral with massless propagators and arbitrary powers governed by the operator scaling dimensions~\cite{Bzowski:2020kfw}.

\subsubsection{Higher-point functions}

Generalising the case for three-point functions, we define the following class of multiple-$K$ conformal integrals:\footnote{Strictly speaking, it is the combination
\begin{equation}
    \tilde{\I}^{(n)}_{\alpha, \bm{\nu}}(p_1, \ldots, p_n) \equiv \p^{\bm{\nu}} \I^{(n)}_{\alpha, \bm{\nu}}(p_1, \ldots, p_n) \,, \quad \text{with} \quad \p^{\bm{\nu}} \equiv \prod_{j=1}^n p_j^{\nu_j} \,,
\end{equation}
rather than the bare triple-$K$ integral $\I^{(n)}_{\alpha, \bm{\nu}}$, that is conformally invariant.
}
\begin{equation}
\label{eq: def conformal multi-K integral}
    \boxed{
    \I^{(n)}_{\alpha, \bm{\nu}}(p_1, \ldots, p_n) \equiv \int_0^\infty \d z z^{\alpha-1} \prod_{j=1}^n K_{\nu_j}(p_j z) \,,
    }
\end{equation}
where $\bm{\nu} \equiv (\nu_1, \ldots, \nu_n)$. In the IR, a sufficient condition for absolute convergence is
\begin{equation}
    \alpha > \sum_{j=1}^n |\Re \, \nu_j| \,.
\end{equation}
Whenever this condition is not satisfied, the integral is understood by analytic continuation, which may be implemented by introducing infinitesimal shifts of the operator dimensions (or equivalently the parameters $\nu_j$) together with the spacetime dimension $d$. The multiple-$K$ conformal integrals~\eqref{eq: def conformal multi-K integral} constitute the main focus of this work. The shadow symmetry $\Delta_j \to d-\Delta_j$, equivalent to $\nu_j \to -\nu_j$ (for $j=1, \ldots, n$), leaves the multiple-$K$ integrals invariant:
\begin{equation}
    \I^{(n)}_{\alpha, \{\nu_1 \cdots \nu_n\}} = \I^{(n)}_{\alpha, \{\sigma_1\nu_1 \cdots \sigma_n\nu_n\}} \,, \quad \text{with} \quad \sigma_j=\pm1\,, \quad (j=1, \ldots, n)\,.
\end{equation}
Additionally, the multiple-$K$ integral is homogeneous:
\begin{equation}
\label{eq: homogeneity}
    \I^{(n)}_{\alpha, \bm{\nu}}(\lambda p_1, \ldots, \lambda p_n) = \lambda^{-\alpha}\I^{(n)}_{\alpha, \bm{\nu}}(p_1, \ldots, p_n) \,.
\end{equation}
Beyond their multiple-$K$ representation, the master integrals~\eqref{eq: def conformal multi-K integral} admit several equivalent formulations. In particular, they can be recast as Feynman integrals over a $(n-1)$-simplex, providing a direct connection with conformal Feynman integrals and contact Witten diagrams in AdS~\cite{Bzowski:2019kwd, Bzowski:2020kfw}.

\paragraph{Example: contact Witten diagrams.} An important application of multiple-$K$ integrals is to contact Witten diagrams. For a non-derivative interaction, the $n$-point contact correlator is obtained by integrating the product of $n$ bulk-to-boundary propagators over the bulk radial coordinate:
\begin{equation}
    \begin{aligned}
        \lla \O_1(\p_1) \cdots \O_n(\p_n) \rra &= \int_0^\infty \frac{\d z}{z^{d+1}} \left(\prod_{j=1}^n  c_{\nu_j}^\AdS z^{\frac{d}{2}}p_j^{\nu_j} K_{\nu_j}(p_j z)\right) \\
        &\propto \p^{\bm{\nu}} \, \I^{(n)}_{d(n/2-1), \bm{\nu}}(p_1, \ldots, p_n)\,,
    \end{aligned}
\end{equation}
where $c_{\nu_j}^\AdS \equiv 2^{1-\nu_j}/\Gamma(\nu_j)$ is the normalisation of the scalar bulk-to-boundary propagator. Upon Wick rotating the radial coordinate according to $z\to e^{\pm i\pi/2} \tau$, where $\tau$ is conformal time and $\pm$ denotes either in-in branch, the AdS contact diagram analytically continues to the corresponding in-in contact diagram in de Sitter spacetime, see~\cite{Bzowski:2023nef, Sleight:2021plv, DiPietro:2021sjt} for details. These correlators form the fundamental building blocks of massive cosmological correlators~\cite{Belrhali:2026ktb, Belrhali:2026rkn, Belrhali:2026act, Grafe:2026qsm, Grafe:2026avi}.

%-------------------------------------
%-------------------------------------
%-------------------------------------
\section{Method of Brackets}
\label{sec: method of brackets}

Having introduced the main objects of interest---multiple-$K$ conformal integrals---, we now turn to the computational framework that will be used throughout this work: the {\it method of brackets}. As we will see, this method essentially reduces the problem of evaluating definite integrals to solving a linear system of algebraic equations. Originally developed in~\cite{gonzalez2008definiteintegralsmethodbrackets, gonzalez2010methodbrackets2examples, Gonzalez:2021vqh, Koutschan2011TheII}, building on earlier developments~\cite{HALLIDAY1987241, Dunne:1987am, Dunne:1987qb}, the method is well suited to the evaluation of Mellin transforms, and has proved to be remarkably effective in a variety of contexts, most notably in the evaluation of Feynman integrals~\cite{GONZALEZ2015214, Gonzalez2017, Ananthanarayan:2021not, Ghosh:2022net}.

\paragraph{Elementary example.} We begin with a simple example illustrating the basic ideas underlying the method. Consider the elementary integral
\begin{equation}
    \I(a) = \int_0^\infty \d z \, e^{-a z} \,, \quad (a>0) \,.
\end{equation}
At first sight, one might attempt what appears to be the worst strategy: expand the exponential in its Taylor series about the origin. Even more surprisingly, we formally interchange the resulting infinite sum with the integral, obtaining a series whose individual terms are all divergent,
\begin{equation}
    \I(a) \stackrel{!}{=} \sum_{n=0}^\infty \frac{(-1)^n}{n!} \, a^n \int_0^\infty \d z \, z^n \,.
\end{equation}
The key idea of the method of brackets is to assign a formal meaning to these divergent integrals. To this end, we introduce the {\it bracket} $\langle b \rangle \equiv \int_0^\infty \d z \, z^{b-1}$ for $b\in\mathbb{C}$, and associate to the original integral the formal {\it bracket series}:
\begin{equation}
    \I(a) \, \widehat{=} \, \sum_n \phi_n a^n \langle n+1 \rangle \,.
\end{equation}
where $\phi_n \equiv \tfrac{(-1)^n}{n!}$. We use the notation ``$\widehat{=}$'' to indicate that this is a formal assignment rather than a rigorous equality. The crucial step now is to solve the {\it bracket equation}: $n+1=0$, whose unique solution is $n^*=-1$. The value of the series is then obtained by evaluating the remaining factors at this solution,
\begin{equation}
    \I(a) = a^{n^*} \Gamma(-n^*) = a^{-1} \Gamma(1) = \frac{1}{a} \,,
\end{equation}
thereby reproducing the exact result. We now formulate this prescription in full generality.

\subsection{General formalism}

The method of brackets is a generalisation of Ramanujan’s master theorem to multiple variables, and is based on a small number of elementary rules.

\subsubsection{Ramanujan’s master theorem}

Generalising the above elementary example, {\it Ramanujan's master theorem} provides a systematic prescription for evaluating the Mellin transform of a function from its Taylor expansion around the origin. At first sight, this procedure is ill-defined, as the resulting term-by-term Mellin transforms are typically divergent. The remarkable content of Ramanujan's theorem is that it provides the appropriate analytic continuation, thereby assigning a finite and well-defined value to these otherwise divergent expressions.

\vskip 4pt
Consider a function $f:z\to f(z)$ of a single variable $z\in\mathbb{C}$, that admits a well-defined Taylor expansion around the origin $z=0$:
\begin{equation}
    f(z) = \sum_{n=0}^\infty a_n \, \frac{(-1)^n}{n!} \, z^n \,,
\end{equation}
with $a_0\neq 0$. Then, under some growth conditions on the coefficients $a_n$ ($n\in\mathbb{N}$), we have
\begin{equation}
    \M[f](s) \equiv \int_0^\infty \d z \, z^{s-1} f(z) = a(-s) \Gamma(s) \,, \quad (\Re\, s>0) \,,
\end{equation}
where $a(-s)$ is interpreted as the natural analytic continuation of the sequence $\{a_n\}_{n\in\mathbb{N}}$ to complex values $s\in\mathbb{C}$. The growth conditions, given by Carlson’s theorem, ensure the {\it uniqueness} of the analytic continuation, see App.~A of~\cite{Raman:2025tsg} for more details.

\paragraph{Practical bracket rules.} Let's now formulate this theorem as a practical computational tool. We define a {\it bracket} as the following divergent integral
\begin{equation}
    \langle a \rangle \equiv \int_0^\infty \d z \, z^{a-1} \,.
\end{equation}
Consider a function $f(z)$ with a general Taylor expansion: $f(z) = \sum_{n=0}^\infty a_n \tfrac{(-1)^n}{n!} z^{\alpha n}$ with $\alpha\in\mathbb{R}$. Then:
\begin{itemize}
    \item {\bf Rule 1.} Assign the following {\it bracket series} to the Mellin transform of the function $f$:
    \begin{equation}
        \M[f](s) = \int_0^\infty \d z \, z^{s-1} f(z) \, \widehat{=} \, \sum_n \phi_n a_n \langle \alpha n+s \rangle \,,
    \end{equation}
    where we define the {\it indicator} of $n$:
    \begin{equation}
        \phi_n \equiv \frac{(-1)^n}{n!} = \underset{z = -n}{\Res} \, \Gamma(z) \,.
    \end{equation}
    
    \item {\bf Rule 2.} The bracket series evaluates to:
    \begin{equation}
        \sum_n \phi_n a_n \langle \alpha n+s \rangle = \frac{1}{|\alpha|} \, a_{-n^*} \Gamma(-n^*) \,,
    \end{equation}
    where $n^*$ is the solution to the {\it bracket equation}: $\alpha n+s=0$.
\end{itemize}

\subsubsection{Multi-dimensional generalisation}
\label{subsubsec: multi-dimensional generalisation}

We now generalise this method to multiple variables. Consider the following multi-dimensional Mellin integral:
\begin{equation}
\label{eq: multi-dim Mellin integral}
    \M[f](\s) \equiv \int_{\mathbb{R}_+^n} \left(\prod_{i=1}^n \d z_i \, z_i^{s_i-1}\right) \, f(\z) \,,
\end{equation}
integrated over the positive orthant, where $\s = (s_1, \ldots, s_n)$ and $\z=(z_1, \ldots, z_n)$, $n\in\mathbb{N}_{>0}$. The method of brackets is based on a small set of heuristic rules:

\begin{itemize}
    \item {\bf Rule 1.} Expand the multivariate function $f$ around the origin in all variables:
    \begin{equation}
    \label{eq: multi-dim Taylor}
        f(\z) = \sum_{\substack{m_i=0\\ i=1,\ldots,n}}^\infty \phi_{\{m\}} \, a_{\{m\}} \left(\prod_{i=j}^n z_j^{\beta_j m_j}\right) \,,
    \end{equation}
    where we define the multi-index indicator:
    \begin{equation}
        \phi_{\{m\}} \equiv \phi_{m_1} \cdots \phi_{m_n} \,,
    \end{equation}
    and $\{m\}$ should be understood as the collection of all indices (similarly for $a_{\{m\}}$).

    \item {\bf Rule 2.} As is often encountered, a multinomial in the integrand is formally replaced by the following bracket series:
    \begin{equation}
    \label{eq: multinomial bracket}
        \frac{\Gamma(s)}{(a_1 + \cdots + a_n)^s} \, \widehat{=} \, \sum_{m_1} \cdots \sum_{m_n} \phi_{\{m\}} \, a_1^{m_1} \cdots a_n^{m_n} \, \langle s + m_1 + \cdots + m_n \rangle \,.
    \end{equation}
    Here, $a_i$ ($i=1, \ldots, n$) can depend on the integration variables $z_i$. This identity can be recovered by using the standard Schwinger parametrisation $\Gamma(s)/(a_1+\cdots+a_n)^s = \int_0^\infty \d z z^{s-1} e^{-(a_1+\cdots+a_n)z}$, and expanding each term $e^{-a_i z}$ in its Taylor series.

    \item {\bf Rule 3.} After expanding the multivariate integrand around the origin in all integration variables $z_i$ ($i=1, \ldots, n$), gathering all powers of $z_i$ together, and using rule 2 to introduce additional bracket series from~\eqref{eq: multinomial bracket}, the integral~\eqref{eq: multi-dim Mellin integral} is assigned the following bracket series:
    \begin{equation}
    \label{eq: multi-dim bracket series}
        \begin{aligned}
            \M[f](\s) \, &\widehat{=} \, \sum_{m_1, \ldots, m_r} \phi_{\{m\}} \, a_{\{m\}} \, \prod_{j=1}^N \langle \beta_{j1}m_1 + \cdots + \beta_{jr}m_r + s_j \rangle \,.
        \end{aligned}
    \end{equation}
    Defining $\bm{B}$ the $N\times r$ matrix with coefficients $(\bm{B})_{ij} \equiv \beta_{ij}$ (with $N\geq n$), $\bm{m} \equiv (m_1, \ldots, m_r)$ and $\bm{s} \equiv (s_1, \ldots, s_N)$, the linear system:
    \begin{equation}
    \label{eq: multi-dim bracket equation}
        \bm{B}\cdot \bm{m} + \s = 0 \,,
    \end{equation}
    is called the {\it bracket equation}.

    \item {\bf Rule 4.} The bracket equation~\eqref{eq: multi-dim bracket equation} is a set of $N$ linear equations in $r$ variables. The difference between the number of sums $r$ and the number of brackets $N$ defines the {\it rank} $R\equiv r-N \geq0$ of the system, i.e.~the number of free indices. Solving the bracket equation leads to several cases:
    \begin{itemize}
        \item[$R=0$:] The bracket equation has a unique solution $\{m^*\}$ and the bracket series~\eqref{eq: multi-dim bracket series} is assigned the following expression:
        \begin{equation}
            \M[f](\s) \, \widehat{=} \, \frac{1}{|\bm{B}|} \, a_{\{m^*\}} \, \Gamma(-m_1^*) \cdots \Gamma(-m_r^*) \,.
        \end{equation}

        \item[$R>0$:] The bracket equation~\eqref{eq: multi-dim bracket equation} does not have a unique solution. We then {\it choose} a subset $\sigma\in \{1, \ldots, r\}$, with $|\sigma| = R$, of free indices $m_i$ with $i\in\sigma$, and denote by $\bar{\sigma}=\{1, \ldots, r\}/\sigma$ the corresponding solutions $m_i^*$ with $i\in\bar{\sigma}$. We should consider all possible choices. For a given choice, the bracket series~\eqref{eq: multi-dim bracket series} is assigned the following $R$-fold series:
        \begin{equation}
            \M[f](\s) \, \widehat{=} \, \frac{1}{|\bm{B}_\sigma|} \, \sum_{i\in\sigma} \phi_{\{m_\sigma\}} \, a_{\{m_{\bar{\sigma}}\}} \left(\prod_{j\in\bar{\sigma}} \Gamma(-m_j^*)\right) \,,
        \end{equation}
        where $\bm{B}_\sigma$ is the submatrix of $\bm{B}$ with columns labelled by $\sigma$ removed. As is often the case in physics, one must sum over the unknown or unconstrained.
    \end{itemize}
\end{itemize}

The final result is determined by summing all convergent series expression found by solving the bracket equation. Divergent series are discarded.

\subsection{Preliminary applications}
\label{subsec: preliminary applications}

As is usually the case, a method is best understood through explicit examples rather than an abstract discussion alone. We therefore illustrate the method of brackets with a series of examples of increasing complexity, some of them already involving Bessel functions.

\subsubsection{Bracket solutions as analytic continuations}
\label{subsubsec: bracket solutions as analytic continuations}

\paragraph{Importance of the phase.} To illustrate the importance of factorising out the indicator $\phi_{\{m\}}$ from the integrand series expansion~\eqref{eq: multi-dim Taylor}, we first use the method of brackets to recover the following known identity:
\begin{equation}
    \I(\nu) \equiv \int_0^\infty \d z \, K_\nu(z) = \frac{\pi}{2\cos(\pi\nu/2)} \,,
\end{equation}
valid for $|\Re(\nu)|<1$ and elsewhere by analytic continuation. To do this, we insert the series expansion around $z=0$ of the Bessel-$K$ function after projecting onto the Bessel-$I$ basis using
\begin{equation}
\label{eq: K/I connection formula and I series}
    K_\nu(z) = \frac{\pi}{2\sin(\pi\nu)} \left[I_{-\nu}(z) - I_{+\nu}(z)\right]\,, \quad I_\nu(z) = \sum_{n=0}^\infty \frac{(z/2)^{2n+\nu}}{n! \Gamma(\nu+n+1)} \,.
\end{equation}
The original integral is therefore given by
\begin{equation}
    \I(\nu) = \I_- - \I_+ \,, \quad \text{with} \quad \I_\pm \, \widehat{=} \, \frac{\pi 2^{\mp\nu-1}}{\sin(\pi\nu)} \sum_{n} \frac{2^{-2n}}{n! \Gamma(n\pm\nu+1)} \, \langle 2n\pm \nu+1\rangle\,,
\end{equation}
where we have integrated term by term and have introduced the bracket. Solving the bracket equation for $\I_-$ gives the unique solution $n^*=(\nu-1)/2$, and evaluating the bracket series on the solution yields
\begin{equation}
    \I_- = \frac{\pi}{2\sin(\pi\nu)} \, e^{+i\pi(\nu-1)/2} \,,
\end{equation}
where, importantly, we have written $1/n! = (-1)^n \phi_n = e^{+i\pi n} \phi_n$ (the sign in the exponential is unimportant). Similarly, the second integral is given by
\begin{equation}
    \I_+ = \frac{\pi}{2\sin(\pi\nu)} \, e^{-i\pi(\nu+1)/2} \,.
\end{equation}
Eventually, collecting both terms, we obtain
\begin{equation}
    \I(\nu) = \frac{\pi}{2} \frac{e^{-i\pi/2}}{\sin(\pi\nu)} \left(e^{+i\pi\nu/2} - e^{-i\pi\nu/2}\right) = \pi \, \frac{\sin(\pi\nu/2)}{\sin(\pi\nu)} \,,
\end{equation}
which equals the original identity after trivial manipulation. A key takeaway from this example is the importance of carefully tracking the phase factors (here $e^{+i\pi n}$) throughout the computation.

\paragraph{Rediscovering the ${}_2F_1$ hypergeometric function.} We now turn to an example in which the bracket equation has non-zero rank. In this case, we show that the different series solutions obtained by choosing different sets of free indices in the bracket series are related by analytic continuation. Consider the following integral:
\begin{equation}
    \I_{\alpha, \nu}(u) \equiv \int_0^\infty \d z z^{\alpha-1} e^{-u z} J_\nu(z) \,, \quad (u>0)\,.
\end{equation}
Expanding both the exponential and the Bessel function using their series expansion around the origin:
\begin{equation}
    e^{-u z} = \sum_{m=0}^\infty \phi_m u^m z^m \,, \quad J_\nu(z) = \sum_{n=0}^\infty \phi_n \, \frac{(z/2)^{2n+\nu}}{\Gamma(\nu+n+1)} \,,
\end{equation}
and integrating term by term, we obtain
\begin{equation}
    \I_{\alpha, \nu}(u) \, \widehat{=} \, \sum_{m, n} \phi_{m, n} \, \frac{u^m}{2^{2n+\nu} \Gamma(\nu+n+1)} \langle m+2n +\nu+\alpha\rangle \,.
\end{equation}
The bracket equation is $m+2n+\nu+\alpha=0$, with rank $1$. Let's first solve for $m^* = -\alpha-\nu-2n$ with $n$ left free. We obtain the one-fold series
\begin{equation}
    \I_{\alpha, \nu}(u) = \frac{1}{2^\nu u^{\nu+\alpha}} \sum_{n=0}^\infty \frac{(-1)^n}{n! 2^{2n}} \frac{\Gamma(2n+\nu+\alpha)}{\Gamma(n+\nu+1)} \frac{1}{u^{2n}} \,.
\end{equation}
This series converges outside the unit disc, $1<|u|$, and can be rewritten in terms of the ${}_2F_1$ hypergeometric function using the duplication formula on the $\Gamma$ factors:
\begin{equation}
    \I_{\alpha, \nu}(u) = \frac{1}{2^\nu u^{\nu+\alpha}} \frac{\Gamma(\nu+\alpha)}{\Gamma(\nu+1)} \, {}_2F_1\left[\left.\begin{matrix} \frac{\nu+\alpha}{2},\,  \frac{\nu+\alpha+1}{2}\\ \nu+1 \end{matrix}\right\vert\frac{-1}{u^2}\right] \,, \quad 1<|u| \,.
\end{equation}
If instead we solve for $n^*=-(m+\nu+\alpha)/2$ with $m$ left as a free index, we obtain
\begin{equation}
    \I_{\alpha, \nu}(u) = 2^{\alpha-1} \sum_{m=0}^\infty \frac{\Gamma(\tfrac{m+\nu+\alpha}{2})}{\Gamma(\nu+1-\tfrac{m+\nu+\alpha}{2})} \frac{(-2u)^m}{m!} \,,
\end{equation}
which, after splitting among even/odd indices, can be written as a sum of ${}_2F_1$ hypergeometric functions:
\begin{equation}
    \begin{aligned}
        \I_{\alpha, \nu}(u) &= \frac{2^{\alpha-1}\Gamma(\tfrac{\nu+\alpha}{2})}{\Gamma(1+\tfrac{\nu-\alpha}{2})}\, {}_2F_1\left[\left.\begin{matrix} \frac{\alpha+\nu}{2},\,  \frac{\alpha-\nu}{2}\\ 1/2 \end{matrix}\right\vert-u^2\right] \\
        &-\frac{2^\alpha\Gamma(\tfrac{\nu+\alpha+1}{2})}{\Gamma(\tfrac{1+\nu-\alpha}{2})} \, u \, {}_2F_1\left[\left.\begin{matrix} \frac{\alpha+\nu+1}{2},\,  \frac{\alpha-\nu+1}{2}\\ 3/2 \end{matrix}\right\vert-u^2\right] \,, \quad |u|<1 \,.
    \end{aligned}
\end{equation}
We therefore have successfully recovered the analytic continuation of the hypergeometric function, the domain of convergence being determined by the index we choose to solve for, equivalently the one we decide to leave free.

\subsubsection{More examples}

For pedagogical purposes, we supplement this section with additional examples, both taken from~\cite{gonzalez2008definiteintegralsmethodbrackets}.

\paragraph{Walli's formula.} Consider the following class of integrals:
\begin{equation}
    \I_m \equiv \int_0^\infty \frac{\d z}{(1+z^2)^{m+1}} \,, \quad (m\in\mathbb{N}_{\geq0}) \,.
\end{equation}
Using the multinomial bracket expansion~\eqref{eq: multinomial bracket}:
\begin{equation}
    \frac{1}{(1+z^2)^{m+1}} = \sum_{n_1, n_2} \phi_{n_1, n_2} \, z^{2n_2} \, \frac{\langle m+1+n_1+n_2 \rangle}{\Gamma(m+1)} \,,
\end{equation}
and integrating term by term, we obtain the following bracket representation
\begin{equation}
    \I_m \, \widehat{=} \, \sum_{n_1, n_2} \phi_{n_1, n_2} \, \frac{\langle m+1+n_1+n_2 \rangle \langle 2n_2+1 \rangle}{\Gamma(m+1)} \,.
\end{equation}
The bracket equations are:
\begin{equation}
    m+1+n_1+n_2=0 \,, \quad 2n_2+1=0 \,,
\end{equation}
which admit the unique solution $n_1^*=-(m+1/2)$ and $n_2^*=-1/2$. Evaluating the bracket series on this solution gives:
\begin{equation}
    \I_m = \frac{1}{2} \frac{\Gamma(-n_1^*)\Gamma(-n_2^*)}{\Gamma(m+1)} = \frac{1}{2} \frac{\Gamma(m+1/2)\Gamma(1/2)}{\Gamma(m+1)} \,,
\end{equation}
which reproduces the expected result.

\paragraph{Multi-dimensional integral.} Finally, let us consider the following class of multi-dimensional integrals:
\begin{equation}
    \I_{s, \bm{p}, \bm{q}}^{(n)}(\bm{r}) \equiv \int_0^\infty \frac{\d z_1 \cdots \d z_n \, z_1^{p_1-1} \cdots z_n^{p_n-1}}{[1+(r_1z_1)^{q_1}+\cdots+(r_nz_n)^{q_n}]^s} \,, \quad (n\in\mathbb{N}_{>0}) \,,
\end{equation}
with $\bm{r}\equiv(r_1, \ldots, r_n)$ and similarly for $\bm{p}$ and $\bm{q}$. We again use the multinomial bracket expansion~\eqref{eq: multinomial bracket} to write the denominator as:
\begin{equation}
    \sum_{m_0, m_1, \ldots, m_n} \phi_{\{m\}} \prod_{j=1}^n (r_j z_j)^{q_j m_j} \, \frac{\langle s+m_0+m_1+\cdots+m_n\rangle}{\Gamma(s)} \,.
\end{equation}
Integrating term by term yields the following bracket representation
\begin{equation}
    \I_{s, \bm{p}, \bm{q}}^{(n)}(\bm{r}) \, \widehat{=} \, \sum_{m_0, m_1, \ldots, m_n} \phi_{\{m\}}  \frac{\langle s+m_0+m_1+\cdots+m_n\rangle}{\Gamma(s)} \prod_{j=1}^n r_j^{q_j m_j} \langle p_j+q_j m_j\rangle \,.
\end{equation}
The bracket equations are a linear system of rank zero (we have $n+1$ indices and $n+1$ equations), whose unique solution is given by
\begin{equation}
    m_j^* = -\frac{p_j}{q_j}\,, \quad (1\leq j \leq n)\,, \quad \text{and} \quad m_0^*=-s+\sum_{j=1}^n \frac{p_j}{q_j} \,.
\end{equation}
Evaluating the bracket series on this solution yields
\begin{equation}
    \I_{s, \bm{p}, \bm{q}}^{(n)}(\bm{r}) = \frac{1}{\Gamma(s)} \Gamma\left(s-\sum_{j=1}^n\frac{p_j}{q_j}\right) \prod_{j=1}^n \frac{\Gamma(p_j/q_j)}{q_j r_j^{p_j}} \,.
\end{equation}
It is remarkable that such elementary rules provide an efficient way to evaluate seemingly complicated integrals.

%-------------------------------------
%-------------------------------------
%-------------------------------------
\section{Exploring Different Representations}
\label{sec: different representations}

Having set up the method of brackets in full generality, we now exploit it to derive a variety of representations for multiple-$K$ integrals. We first recover the standard representation of the triple-$K$ integral in terms of the generalised Appell $F_4$ hypergeometric series in two variables, whose domain of convergence lies outside the physical region defined by the triangle inequalities. We then rederive this representation by applying a weight-shifting operator to a conformally coupled seed integral, for which we obtain a representation that covers the physical region upon resummation. Finally, we use the method of brackets in conjunction with alternative representations of the Bessel-$K$ function to derive new representations of multiple-$K$ integrals.

\subsection{Triple-$K$ integrals: recovering Appell $F_4$}
\label{subsec: triple-K integrals: recovering Appell F4}

Our goal here is modest but instructive. We show that the bracket formalism reproduces, from scratch, the well-known fact that the momentum-space three-point function of scalar primaries is given by a linear combination of four Appell $F_4$ functions~\cite{Bzowski:2013sza, Coriano:2013jba}, with a series that converges only around the origin $p_1=p_2=0$. We then revisit the computation using a simpler ``seed'' integral, replacing one leg by its conformally coupled counterpart, $K_{1/2}(z)\propto e^{-z}$. Applying the bracket method to this two-Bessel integral and resumming one series yields a representation convergent throughout the physical region. Finally, we relate the seed to the full triple-$K$ integral through an integral transform.
 
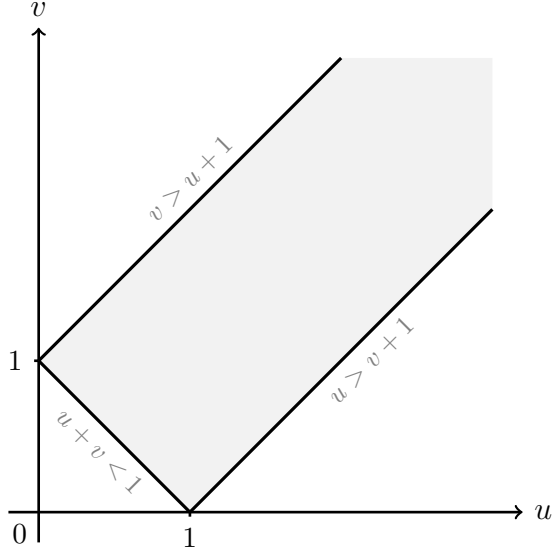
\begin{figure}[h!]
\centering
    \begin{tikzpicture}[line width=1. pt, scale=2]

    % Axes
    \draw[->] (-0.2,0) -- (3.2,0) node[right] {$u$};
    \draw[->] (0,-0.2) -- (0, 3.2) node[above] {$v$};

    % Ticks and labels on x-axis
    \foreach \x in {1} {
        \draw (\x, 0.03) -- (\x, -0.03) node[below] {\small $\x$};
    }
    \node[below left] at (0,0) {\small $0$};
    
    % Ticks and labels on y-axis
    \foreach \y in {1} {
        \draw (0.03, \y) -- (-0.03, \y) node[left] {\small $\y$};
    }

    \fill[black!5] (0,1) -- (2,3) -- (3,3) -- (3,2) -- (1, 0) -- cycle;

    \draw[black, line width=1.2pt] (1,0) -- (0,1);
    \node[gray, rotate=-45] at (0.4, 0.4) {\footnotesize $u+v<1$};

    \draw[black, line width=1.2pt] (0,1) -- (2,3);
    \node[gray, rotate=45] at (1, 2.2) {\footnotesize $v>u+1$};

    \draw[black, line width=1.2pt] (1,0) -- (3,2);
    \node[gray, rotate=45] at (2.2, 1) {\footnotesize $u>v+1$};

    \end{tikzpicture}
    \caption{Kinematic space $(u, v) \equiv (p_1/p_3, p_2/p_3)$ of the triple-$K$ integral, in the positive quadrant. The origin is the double-soft limit $u, v\to0$, and the gray unbounded wedge is the physical region.}
    \label{fig: u/v kinematic regions}
\end{figure}

\paragraph{Physical region.} Before diving into the computation, it is worth recalling what is the physical region for this integral. The momenta $p_i = |\p_i|$ ($i=1, 2, 3$) are the magnitudes of three Euclidean vectors constrained by momentum conservation, $\p_1 + \p_2 + \p_3 = 0$. Consequently they are not independent: they are the side lengths of a (possibly degenerate) triangle, and must satisfy the three triangle inequalities
\begin{equation}
\label{eq: triangle inequalities}
    p_1 < p_2 + p_3\,, \qquad p_2 < p_1 + p_3\,, \qquad p_3 < p_1 + p_2\,.
\end{equation}
In terms of the ratios 
\begin{equation}
    u \equiv \frac{p_1}{p_3}\,, \quad v\equiv \frac{p_2}{p_3} \,,
\end{equation}
with $u, v>0$, it maps to the region $u+v>1$, $|u-v|<1$ in the $(u,v)$ plane---an unbounded wedge that entirely misses the simplex around the origin, see Fig.~\ref{fig: u/v kinematic regions}. As we will see momentarily, the series solutions obtained directly from the small-$p_1,p_2$ expansion of the Bessel functions converge only inside that simplex, and are therefore of no direct use in the physical region~\eqref{eq: triangle inequalities}.

\subsubsection{Series from brackets}

We first deal with $\I^{(3)}_{\alpha,\bm\nu}$ head on by expanding each Bessel-$K$ factor around the origin. Since $K_\nu$ itself does not admit a single-valued Taylor series when $\nu\notin\mathbb Z$, we use the standard connection formula to the Bessel-$I$ function~\eqref{eq: K/I connection formula and I series}, written in the following compact form:
\begin{equation}
    K_\nu(z) = \frac{\pi}{2\sin\pi\nu}\sum_{\sigma=\pm}(-\sigma) I_{\sigma\nu}(z) \,.
\end{equation}
Applying this to each of the three legs produces $2^3=8$ terms, labelled by signs $(\sigma_1, \sigma_2, \sigma_3)=\pm$:
\begin{equation}
    \I^{(3)}_{\alpha, \bm{\nu}} = \left[\prod_{j=1}^3 \frac{\pi}{2\sin(\pi\nu_j)}\right] \sum_{\sigma_1, \sigma_2, \sigma_3=\pm} (-\sigma_1\sigma_2\sigma_3)\, \I_{\sigma_1\sigma_2\sigma_3} \,,
\end{equation}
where each $\I_{\sigma_1\sigma_2\sigma_3}$ is obtained by inserting the series~\eqref{eq: K/I connection formula and I series} for $\I_{\sigma_j\nu_j}(p_jz)$ ($j=1, 2, 3$) and integrating term by term. This is a direct application of Rule 3 of Sec.~\ref{subsubsec: multi-dimensional generalisation}:
three summation indices $n_1,n_2,n_3$ and a single bracket coming from the
$z$-integral,
\begin{equation}
    \I_{\sigma_1\sigma_2\sigma_3} \, \widehat{=} \, \sum_{n_1, n_2, n_3=0}^\infty \phi_{\{n\}}\, e^{i\pi n_{123}} \left[\prod_{j=1}^3 \frac{(p_j/2)^{2n_j+\sigma_j\nu_j}}{\Gamma(\sigma_j\nu_j+n_j+1)}\right] \, \langle 2n_{123} + \nu_{\sigma_1\sigma_2\sigma_3}+\alpha \rangle \,,
\end{equation}
where $\nu_{\sigma_1\sigma_2\sigma_3}\equiv\sigma_1\nu_1+\sigma_2\nu_2+\sigma_3\nu_3$ and the phase $e^{i\pi n_{123}}$ keeps track of the branch of $1/n_j!$ as in the elementary example of Sec.~\ref{subsubsec: bracket solutions as analytic continuations}. With three indices and a single bracket equation, the rank of the system is two, so we must leave two indices free. Solving for $n_3^\ast = -n_{12}-(\nu_{\sigma_1\sigma_2\sigma_3}+\alpha)/2$ and resumming the resulting double series in terms of Pochhammer symbols, one finds a two-variable hypergeometric series that is recognised as an Appell $F_4$ function,
\begin{equation}
    \begin{aligned}
        \I_{\sigma_1\sigma_2\sigma_3} &= \frac{2^{\alpha-1}}{p_3^\alpha} \frac{\Gamma\big(\tfrac{\nu_{\sigma_1\sigma_2\sigma_3}+\alpha}{2}\big)}{\Gamma\big(1-\tfrac{\sigma_1\nu_1+\sigma_2\nu_2-\sigma_3\nu_3+\alpha}{2}\big)\Gamma(\sigma_1\nu_1+1)\Gamma(\sigma_2\nu_2+1)} \\
        &\times u^{\sigma_1\nu_1} v^{\sigma_2\nu_2}\,
        F_4\!\left[\begin{matrix} \tfrac{\sigma_1\nu_1+\sigma_2\nu_2+\sigma_3\nu_3+\alpha}{2},\; \tfrac{\sigma_1\nu_1+\sigma_2\nu_2-\sigma_3\nu_3+\alpha}{2}\\[2pt] \sigma_1\nu_1+1,\; \sigma_2\nu_2+1 \end{matrix}\,\middle|\, u^2, v^2 \right] \,,
    \end{aligned}
\end{equation}
with $u \equiv p_1/p_3$, $v\equiv p_2/p_3$. This series converges for $|u|+|v|<1$. Solving instead for $n_1^\ast$ or $n_2^\ast$ produces the two familiar alternative branches, convergent respectively for $1+v<u$ or $1+u<v$: exactly the analytic continuations of $F_4$ across the boundaries of its natural domain, see Fig.~\ref{fig: u/v kinematic regions}, in complete analogy with the ${}_2F_1$ example of Sec.~\ref{subsubsec: bracket solutions as analytic continuations}. None of these three branches, however, covers the physical wedge~\eqref{eq: triangle inequalities}.

\vskip 4pt
At this stage, summing the $2^3=8$ terms $\I_{\sigma_1\sigma_2\sigma_3}$ naively would leave the somewhat unwieldy triple sum over $(\sigma_1, \sigma_2, \sigma_3)$. It collapses to only four terms once one notices that $F_4(a,b;c_1,c_2\,|\,x,y)$ is symmetric under $a\leftrightarrow b$: writing $\beta_1\equiv\sigma_1\nu_1$, $\beta_2\equiv\sigma_2\nu_2$, the two hypergeometric parameters above are $a_{\sigma_3} = \tfrac{\alpha+\beta_{12}+\sigma_3\nu_3}{2}$ and $b_{\sigma_3}=\tfrac{\alpha+\beta_{12}-\sigma_3\nu_3}{2}$, which are simply exchanged, $a_+ = b_-$ and $b_+=a_-$, when $\sigma_3\to-\sigma_3$. The two terms $\sigma_3=\pm$ therefore multiply the \emph{same} $F_4$ function, and only their Gamma-function prefactors need to be combined. Carefully tracking the phases $e^{\pm i\pi n_j}$ inherited from~\eqref{eq: K/I connection formula and I series}---exactly as emphasised in the introductory example of Sec.~\ref{subsubsec: bracket solutions as analytic continuations}---the reflection formula $\Gamma(z)\Gamma(1-z)=\pi/\sin\pi z$ applied simultaneously to the $\nu_1$-, $\nu_2$- and $\nu_3$-dependent Gamma functions turns the sum over $\sigma_3$ into a single, purely Gamma-function prefactor, and the overall $\prod_{j=1}^3 \pi/(2\sin\pi\nu_j)$ cancels completely. What remains is a sum over only the four sign choices $\bm\beta=\pm\bm\nu\equiv(\pm\nu_1,\pm\nu_2)$:
\begin{equation}
\label{eq: triple-K F4 series rep}
    \boxed{
    \begin{aligned}
        \I_{\alpha, \bm{\nu}}^{(3)} = \frac{2^{\alpha-4}}{p_3^\alpha} \sum_{\bm{\beta}=\pm\bm{\nu}} &\Gamma\left[\frac{\alpha+\beta_{12}+\nu_3}{2}, \frac{\alpha+\beta_{12}-\nu_3}{2}, -\beta_1, -\beta_2 \right] \\
        &\times u^{\beta_1} v^{\beta_2} \, 
        F_4 \left[\begin{matrix}
        \frac{\alpha+\beta_{12}+\nu_3}{2},\,\, \frac{\alpha+\beta_{12}-\nu_3}{2} \\[2pt] 1+\beta_1, \,\, 1+\beta_2
    \end{matrix}\,\middle|\, u^2, v^2\right] \,.
    \end{aligned}
    }
\end{equation}
This reproduces the known Appell $F_4$ representation of the triple-$K$ integral \cite{Bzowski:2013sza}. As anticipated, the four hypergeometric series in~\eqref{eq: triple-K F4 series rep} converge only inside the simplex $|u|+|v|<1$, well short of the physical wedge~\eqref{eq: triangle inequalities}: this is precisely what motivates the alternative route we explore next.

\subsubsection{Seed integral \& weight-shifting operator}

The derivation above treats all three Bessel-$K$ legs on the same footing, which is economical but, as we just saw, produces a representation with poor convergence properties. As a first step towards a more useful representation---to be developed fully in Sec.~\ref{sec: analytic continuation}---it is instructive to simplify the problem by setting the third leg to be conformally coupled, $K_{\nu_3}(z)\big|_{\nu_3=1/2} \propto e^{-z}$. This defines the \emph{seed integral}
\begin{equation}
\label{eq: seed integral}
    \S^{(2)}_{\alpha, \bm{\nu}} (u, v) \equiv \int_0^\infty \d z\, z^{\alpha-1} e^{-z} K_{\nu_1}(u z) K_{\nu_2}(v z) \,.
\end{equation}
As we show below, its bracket solution can subsequently be converted, by an integral transform, into a solution for the full triple-$K$ integral with generic $\nu_3$.
 
\paragraph{Bracket solution.} We use the integral representation of the Euler-type, given by (\cite[Eq.~(10.38.8)]{NIST} after change of variables $s\to t=s-1 \to t=2s$)
\begin{equation}
\label{eq: Euler-type integral rep}
    K_\nu(z) = \frac{\sqrt\pi\, (2z)^\nu}{\Gamma(\nu+\tfrac12)} \int_0^\infty \d s\, [s(s+1)]^{\nu-\frac12}\, e^{-z(2s+1)} \,,
\end{equation}
for each of the two Bessel factors in~\eqref{eq: seed integral}. The $z$-integral can then be performed explicitly,
\begin{equation}
    \begin{aligned}
        \S^{(2)}_{\alpha,\bm\nu}(u,v) = \frac{\pi (2u)^{\nu_1}(2v)^{\nu_2}}{\Gamma[\tfrac12+\nu_1,\tfrac12+\nu_2]} &\int_0^\infty \d s_1 \d s_2\, [s_1(s_1+1)]^{\nu_1-\frac12} [s_2(s_2+1)]^{\nu_2-\frac12} \\
        &\times\frac{\Gamma(\alpha+\nu_{12})}{[1+u+v+2us_1+2vs_2]^{\alpha+\nu_{12}}}\,,
    \end{aligned}
\end{equation}
which is now a rational function of the two auxiliary Schwinger parameters. Applying
Rule 2 of Sec.~\ref{subsubsec: multi-dimensional generalisation} to the denominator, and the bracket representation of the binomial series to each factor $[s_i(s_i+1)]^{\nu_i-1/2}$ ($i=1, 2$), one
obtains a bracket series with seven indices $\{a_i,b_i\}_{i=1,2}$ and $c_0,c_1,c_2$,
subject to a system of five bracket equations:
\begin{equation}
    \begin{aligned}
        \I_{\alpha, \bm{\nu}} (u, v) \, \widehat{=} \, \frac{\pi}{\Gamma[\tfrac{1}{2}\pm\nu_1, \tfrac{1}{2}\pm\nu_2]} &\sum_{\{a, b, c\}} \phi_{\{a, b, c\}} (1+u+v)^{c_0} (2u)^{c_1+\nu_1} (2v)^{c_2+\nu_2} \\
        & \langle \tfrac{1}{2}-\nu_1+a_1+b_1 \rangle \langle \tfrac{1}{2}-\nu_2+a_2+b_2 \rangle \\
        & \langle \tfrac{1}{2}+\nu_1+a_1+c_1 \rangle \langle \tfrac{1}{2}+\nu_2+a_2+c_2 \rangle \\
        & \langle \alpha+\nu_{12}+c_0+c_1+c_2 \rangle \,,
    \end{aligned}
\end{equation}
where we use the short notation for the product of $\Gamma$-function, $\Gamma[\pm a]\equiv \Gamma(+a)\Gamma(-a)$. We choose to solve the bracket equations by letting either $b_1, b_2=m_1, m_2\in\mathbb{N}$ or $c_1, c_2=m_1, m_2\in\mathbb{N}$ be free indices, see Sec.~\ref{subsubsec: from Euler-type integral} for the generalisation to multiple-$K$ integrals with $n\geq3$ legs. Applying the duplication formula to simplify the resulting Gamma functions, one finds four branches, labelled again by $\bm\beta=\pm\bm\nu$,
\begin{equation}
\label{eq: seed double series}
    \begin{aligned}
        \S^{(2)}_{\alpha, \bm{\nu}} (u, v) &= \tfrac{1}{4}(1+u+v)^{-\alpha} \sum_{\bm{\beta}=\pm \bm{\nu}}\, (x/4)^{-\beta_1} (y/4)^{-\beta_2}\, \Gamma[\beta_1, \beta_2] \\
        &\sum_{m_1, m_2=0}^\infty \Gamma[\alpha-\beta_{12}+m_{12}]\, \frac{(\tfrac{1}{2}-\beta_1)_{m_1}(\tfrac{1}{2}-\beta_2)_{m_2}}{(1-2\beta_1)_{m_1} (1-2\beta_2)_{m_2}} \, \frac{x^{m_1}}{m_1!} \frac{y^{m_2}}{m_2!} \,,
    \end{aligned}
\end{equation}
where we have introduced the new kinematic variables
\begin{equation}
\label{eq: xy variables}
    x \equiv \frac{2u}{1+u+v}\,, \qquad y \equiv \frac{2v}{1+u+v}\,,
\end{equation}
which arise naturally in the course of the derivation. The inverse map is given by: $u=x/(2-x-y)$ and $v=y/(2-x-y)$, and the physical region is mapped to the upper-right triangular region of the unit square:
\begin{equation}
    \{(x, y) \in (0, 1)^2 \, | \, x+y>1\} \,.
\end{equation}
The three boundaries $x=1$, $y=1$ and $x+y=1$ correspond to the folded kinematic configurations, i.e.~degenerate triangles. Each of the four terms in~\eqref{eq: seed double series} is, on its own, precisely the double power series of an Appell $F_2$ function.
 
\paragraph{Resumming one layer.} The double series \eqref{eq: seed double series} is already a considerable improvement, since it is expressed in terms of the bounded variables $x,y\in(0,1)$ rather than the unbounded $u,v$. We can do better: writing $\Gamma[\alpha-\beta_{12}+m_{12}] = \Gamma(\alpha-\beta_{12}+m_1)\, (\alpha-\beta_{12}+m_1)_{m_2}$---a choice that breaks the $1\leftrightarrow 2$ symmetry manifestly, though not in substance---the sum over $m_2$ is recognised as the Taylor series of a Gauss hypergeometric function, and can be resummed in closed form. This leaves a single residual sum over $m$, dressing a ${}_2F_1$:
\begin{equation}
\label{eq: seed convergent series}
    \boxed{
    \begin{aligned}
        \S^{(2)}_{\alpha, \bm{\nu}} &(u, v) = \tfrac{1}{4} (1+u+v)^{-\alpha}\sum_{\bm{\beta}=\pm \bm{\nu}} (x/4)^{-\beta_1} (y/4)^{-\beta_2} \Gamma[\beta_1, \beta_2]\\
        &\times\sum_{m=0}^\infty \frac{\Gamma[\alpha-\beta_{12}+m](\tfrac{1}{2}-\beta_1)_{m}}{(1-2\beta_1)_{m}} \frac{x^{m}}{m!}\;  {}_2F_1\!\left[\begin{matrix} \alpha-\beta_{12}+m,\,  \tfrac{1}{2}-\beta_2\\ 1-2\beta_2 \end{matrix}\,\middle\vert\, y\right] \,.
    \end{aligned}
    }
\end{equation}
This is our first genuinely convergent representation.
 
\paragraph{Weight-shifting operator.} We now close the loop and use the seed integral to re-derive the triple-$K$ result~\eqref{eq: triple-K F4 series rep} directly. The idea is simple: instead of expanding
$K_{\nu_3}(z)$ around the origin, we use its integral representation
\begin{equation}
    K_{\nu_3}(z) = \frac{\sqrt\pi\, 2^{-\nu_3}}{\Gamma(\nu_3+\tfrac12)}\, z^{-\nu_3} \int_1^\infty \d t\, (t^2-1)^{\nu_3-\frac12}\, e^{-zt}\,,
\end{equation}
insert it into the triple-$K$ integral, and perform the $z$-integral at fixed $t$. Rescaling $z\to z/t$, the remaining $z$-integral is recognised as the seed integral~\eqref{eq: seed integral}, evaluated at rescaled arguments and shifted weight $\alpha\to\alpha+\nu_3$:
\begin{equation}
\label{eq: weight shifting relation}
    \I^{(3)}_{\alpha, \bm{\nu}} = \frac{\sqrt\pi\, 2^{-\nu_3}}{\Gamma(\nu_3+\tfrac12)} \int_1^\infty \d t\, \frac{(t^2-1)^{\nu_3-\frac12}}{t^{\alpha+\nu_3}}\, \S^{(2)}_{\alpha+\nu_3, \bm{\nu}}\!\left(\frac{u}{t}, \frac{v}{t}\right) \,.
\end{equation}
This is a weight-shifting relation in disguise: it trades the third Bessel-$K$ leg, with its own parameter $\nu_3$, for an integral kernel acting on the seed integral evaluated at shifted weight $\alpha+\nu_3$. To check consistency with~\eqref{eq: triple-K F4 series rep}, we insert into~\eqref{eq: weight shifting relation} not the resummed form~\eqref{eq: seed convergent series}, but the Appell $F_4$
representation of the seed integral itself,
\begin{equation}
\label{eq: conformal seed F4 series rep}
    \S_{\alpha, \bm{\nu}}^{(2)}(u,v) = \frac{1}{4} \sum_{\bm{\beta}=\pm\bm{\nu}} \left(\frac{u}{2}\right)^{\beta_1} \left(\frac{v}{2}\right)^{\beta_2} \, \Gamma\left[\alpha+\beta_{12}, -\beta_1, -\beta_2 \right]
        F_4 \left[\begin{matrix}
        \frac{\alpha+\beta_{12}}{2},\,\, \frac{\alpha+1+\beta_{12}}{2} \\[2pt] 1+\beta_1, \,\, 1+\beta_2
    \end{matrix}\,\middle\vert\, u^2, v^2\right] \,,
\end{equation}
and expand both $F_4$ series in~\eqref{eq: conformal seed F4 series rep} as double power series in
$m,n$. The $t$-integral in~\eqref{eq: weight shifting relation} then reduces, term by term, to an elementary Beta-type integral,
\begin{equation}
    \int_1^\infty \d t\, \frac{(t^2-1)^{\nu_3-\frac12}}{t^{\alpha+\nu_3+2m+2n+\beta_{12}}} = \frac{\Gamma\!\big(\tfrac{\alpha-\nu_3+\beta_{12}}{2}+m+n\big)\, \Gamma(\nu_3+\tfrac12)}{\Gamma\!\big(\tfrac{\alpha+\nu_3+\beta_{12}+1}{2}+m+n\big)}\,.
\end{equation}
Rearranging the resulting Pochhammer symbols and applying the Legendre duplication formula once more, the double series recombines exactly into the Appell $F_4$ function of~\eqref{eq: triple-K F4 series rep}. 

\subsection{Multiple-$K$ integrals: Lauricella representations}

We now turn to multiple-$K$ integrals, for which the method of brackets naturally produces two complementary representations, of Lauricella type $F_C$ and $F_A$, expressed in terms of the dimensionless kinematic variables it singles out.

\subsubsection{From Schl\"afi integral}
\label{subsubsec: from schlafi series}

We first generalise the derivation of the triple-$K$ integral in Sec.~\ref{subsec: triple-K integrals: recovering Appell F4} to the case of $n$-$K$ integrals with $n\geq3$. To do so, we use Schl\"afi's integral representation, given by~\cite[Eq.~(10.32.10)]{NIST}:
\begin{equation}
    K_\nu(z) = \frac{1}{2}\left(\frac{z}{2}\right)^\nu \int_0^\infty \frac{\d t}{t^{\nu+1}} e^{-t-\frac{z^2}{4t}} \,,
\end{equation}
for each Bessel-$K$ function in the multiple-$K$ integral~\eqref{eq: def conformal multi-K integral}. We obtain
\begin{equation}
    \I_{\alpha, \bm{\nu}}^{(n)} = \frac{1}{2^n} \left(\frac{\p}{2}\right)^{\bm{\nu}} \int_0^\infty\left[\prod_{j=1}^n\frac{\d t_j}{t_j^{\nu_j+1}}\right] e^{-|\bm{t}|} \int_0^\infty \d z z^{\alpha+|\bm{\nu}|-1} e^{-\frac{z^2}{4}\left(\frac{p_1^2}{t_1}+\cdots+\frac{p_n^2}{t_n}\right)} \,,
\end{equation}
where we use multi-index vectors to make notations compact: we introduce $\p\equiv (p_1, \ldots, p_n)$ (similarly for $\bm{t}$ and $\bm{\nu}$), and use the scalar shorthand $|\bm{t}| \equiv \sum_{j=1}^n t_j$ (similarly for $|\bm{\nu}|$). For component-wise products, it should be understood that,~e.g., $(\p/2)^{\bm{\nu}} \equiv \prod_{j=1}^n (p_j/2)^{\nu_j}$. We now expand the exponentials into brackets:
\begin{equation}
    e^{-t_j} \, \widehat{=} \, \sum_{a_j} \phi_{a_j} t_j^{a_j}\,, \quad (j=1, \ldots, n)\,.
\end{equation}
The $z$-integral can be performed explicitly and expanded in brackets applying again Rule 2 of~\ref{subsubsec: multi-dimensional generalisation} to the denominator:
\begin{equation}
    \begin{aligned}
        \int_0^\infty \d z z^{\alpha+|\bm{\nu}|-1} e^{-\frac{z^2}{4}\left(\frac{p_1^2}{t_1}+\cdots+\frac{p_n^2}{t_n}\right)} &= \frac{2^{\alpha+|\bm{\nu}|-1}\Gamma(\tfrac{\alpha+|\bm{\nu}|}{2})}{\left(\frac{p_1^2}{t_1}+\cdots+\frac{p_n^2}{t_n}\right)^{\frac{\alpha+|\bm{\nu}|}{2}}} \\
        &\, \widehat{=} \, 2^{\alpha+|\bm{\nu}|-1} \sum_{b_1, \ldots, b_n} \phi_{\{b\}} \left(\frac{\p^2}{\bm{t}}\right)^{\bm{b}} \langle \tfrac{\alpha+|\bm{\nu}|}{2}+|\bm{b}| \rangle \,.
    \end{aligned}
\end{equation}
The $t_j$-integrals ($j=1, \ldots, n$) introduce $n$ brackets. Eventually, we obtain the following bracket representation for the multiple-$K$ integral:
\begin{equation}
\label{eq: Schlafi bracket representation}
    \I_{\alpha, \bm{\nu}}^{(n)} \, \widehat{=} \, \frac{2^{\alpha+|\bm{\nu}|-1}}{2^n} \left(\frac{\p}{2}\right)^{\bm{\nu}} \sum_{\{a, b\}} \phi_{\{a, b\}} \, \p^{2\bm{b}}\,  \langle \tfrac{\alpha+|\bm{\nu}|}{2} + |\bm{b}| \rangle \langle \bm{a} - \bm{b} - \bm{\nu} \rangle \,.
\end{equation}
The bracket equations:
\begin{equation}
    \tfrac{\alpha+|\bm{\nu}|}{2} + |\bm{b}| = 0\,, \quad a_j-b_j-\nu_j=0\,,\quad (j=1, \ldots, n)\,,
\end{equation}
are a linear system of rank $n-1$ (we have $2n$ indices and $n+1$ equations). We choose to set the exponent of $p_n$ to be negative, therefore picking this momentum to normalise all other momenta $p_j$ for $j=1, \ldots, n-1$. For a given $j=1, \ldots, n-1$, we can then either solve for $a_j$ or $b_j$. This leads to $2^{n-1}$ possible solutions, which should be all considered, and summed over. For example, choosing $b_j=m_j\in\mathbb{N}$ for $j=1, \ldots, n-1$ to be the free indices, the bracket equations imply:
\begin{equation}
    \begin{aligned}
        &b_n = -m_1 - \cdots - m_{n-1} - \frac{\alpha+|\bm{\nu}|}{2} \,, \quad a_n = -m_1 - \cdots - m_{n-1} - \frac{\alpha-|\bm{\nu}|}{2} \,, \\
        &a_j = b_j+\nu_j = m_j + \nu_j\,, \quad (j=1, \ldots, n-1) \,.
    \end{aligned}
\end{equation}
Considering all possible solutions and evaluating the bracket expansion~\eqref{eq: Schlafi bracket representation} on these solutions yields the following $2^{n-1}$ combinations of $n-1$-fold series:
\begin{equation}
    \begin{aligned}
        \I_{\alpha, \bm{\nu}}^{(n)} = \frac{2^{\alpha-n-1}}{p_n^\alpha} \sum_{\bm{\beta}=\pm \bm{\nu}}\sum_{\bm{m}\geq0} \phi_{\{m\}} \left(\frac{\p}{p_n}\right)^{2\bm{m}+\bm{\beta}}\Gamma\left[|\bm{m}|+\frac{\alpha+|\bm{\beta}|\pm \nu_n}{2}\right] \Gamma[-\bm{m}-\bm{\nu}] \,,
    \end{aligned}
\end{equation}
where, here, the multi-index vectors have size $n-1$: $\bm{\nu}\equiv (\nu_1, \ldots, \nu_{n-1})$, $\p\equiv (p_1, \ldots, p_{n-1})$ and $\bm{m} \equiv (m_1, \ldots, m_{n-1})$. Notice that the different solutions are the usual positive and negative frequency branches and summing over them restores the shadow symmetry $\nu_j \leftrightarrow -\nu_j$. Eventually, using Euler's reflection formula and introducing Pochhammer symbols, we recognise the generalised type-$C$ Lauricella function of $n-1$ variables:
\begin{equation}
\label{eq: Lauricella F_C rep}
    \boxed{
    \begin{aligned}
        \I_{\alpha, \bm{\nu}}^{(n)} = \frac{2^{\alpha-n-1}}{p_n^\alpha} \sum_{\bm{\beta}=\pm \bm{\nu}} &\Gamma\left[\frac{\alpha+|\bm{\beta}|\pm \nu_n}{2}\right] \Gamma[-\bm{\beta}]\,\, \bm{u}^{\bm{\beta}} \\
        &\times F_C^{(n-1)} \left[\left.\begin{matrix}
        \frac{\alpha+|\bm{\beta}|+\nu_n}{2}\, ,\, \frac{\alpha+|\bm{\beta}|-\nu_n}{2} \\ 1+\bm{\beta}
    \end{matrix}\right\vert \bm{u}^2 \right]\,,
    \end{aligned}
    }
\end{equation}
where we have defined the natural dimensionless kinematic ratios $\bm{u}\equiv (u_1, \ldots, u_{n-1})$ with
\begin{equation}
    u_j \equiv \frac{p_j}{p_n} \,, \quad (j=1, \ldots, n-1)\,.
\end{equation}
In these variables, the physical region is~\cite{Grafe:2026avi}:
\begin{equation}
    u_i \leq 1+\sum_{j\neq i} u_j \,, \quad (i=1, \ldots, n-1)\,, \quad \text{and} \quad 1\leq \sum_{i=1}^{n-1} u_i \,,
\end{equation}
which generalises the case for $n=3$ in Sec.~\ref{subsec: triple-K integrals: recovering Appell F4}, see Fig.~\ref{fig: u/v kinematic regions}. The choice of $n-1$ indices we solve for in the bracket equations determines the reference momentum. The derived series representation~\eqref{eq: Lauricella F_C rep} only converges in the simplex around the origin, $\sum_{j=1}^{n-1}|u_j|<1$, and does not cover the physical region. Further details about type-$C$ Lauricella functions can be found in App.~A of~\cite{Grafe:2026avi}.

\subsubsection{From Euler-type integral}
\label{subsubsec: from Euler-type integral}

Starting from a different integral representation of the Bessel-$K$ function, we can derive an alternative series representation that converges in a different region of the kinematic space. We consider the integral representation of the Euler-type~\eqref{eq: Euler-type integral rep}. After plugging~\eqref{eq: Euler-type integral rep} into~\eqref{eq: def conformal multi-K integral}, we perform the $z$-integral using the definition of the $\Gamma$-function and expand the result using Rule 2 in~\eqref{eq: multinomial bracket}:
\begin{align}
    \int_0^\infty \d z\, z^{\alpha+|\bm\nu|-1} &e^{-z(S_n+2\bm{p}\cdot\bm{s})} = \frac{\Gamma(\alpha +|\bm\nu|)}{\left(S_n+2\bm{p}\cdot\bm{s}\right)^{\alpha+|\bm\nu|}} \nonumber \\
    &\, \widehat{=} \, \sum\limits_{c_0,c_1,\dots,c_n}\phi_{\{c_0,\bm{c}\}} \,S_n^{c_0} \left[\prod_{j=1}^n(2p_js_j)^{c_j}\right] \langle \alpha+|\bm\nu|+c_0+|\bm{c}|\rangle\,.
\end{align}
Here, the total energy is given by $S_n\equiv p_1+\dots+p_n$, and bold quantities denote vectors of dimension $n$,~i.e.~$\p = (p_1,\dots,p_n)$, $\bm{s} = (s_1,\dots,s_n)$ and $\bm{c}=(c_1,\dots,c_n)$. We also use the shorthand $|\bm\nu| = \sum_{j=1}^n\nu_j$ again. Moreover, we expand the terms
\begin{equation}
    (1+s_j)^{\nu_j-\frac12} \, \widehat{=} \, \frac{1}{\Gamma(\frac12-\nu_j)} \sum\limits_{a_j,b_j} \phi_{a_j, b_j}\, s_j^{a_j} \langle a_j+b_j +\tfrac12-\nu_j\rangle\,, \quad (j=1, \ldots, n)\,,
\end{equation}
in the Euler representation in terms of brackets and then solve the integrals over $s_j$, leading to one more bracket per integration variable. Combining everything, we arrive at the bracket expansion
\begin{equation}
\label{eq: bracket rep from Euler}
    \begin{aligned}
        \I^{(n)}_{\alpha, \bm{\nu}} \, \widehat{=} \, \frac{\pi^{n/2}}{\Gamma(\tfrac12 \pm \bm\nu)} \sum\limits_{\{a,b,c\}} \phi_{\{a,b,c\}} \,S_n^{c_0} (2\bm{p})^{\bm{c}+\bm\nu} &\langle \alpha+|\bm\nu| +c_0+|\bm{c}|\rangle \\
        &\times\langle \bm{a} +\bm{b} +\tfrac12 -\bm\nu\rangle \langle \bm{a} +\bm{c} +\tfrac12 +\bm\nu\rangle\,.
    \end{aligned}
\end{equation}
Solving the bracket equations
\begin{equation}
    \alpha+|\bm\nu| +c_0+|\bm{c}| =0\,, \quad a_j +b_j +\tfrac12 -\nu_j =0\,, \quad
    a_j+c_j +\tfrac12 +\nu_j =0\,,
\end{equation}
for $j=1, \ldots, n$ leads to series representations that---depending on the choice of solution---converge in different kinematic regions. The last two equations are \textit{local}, i.e.~they only relate parameters with the same index $j$. The first equation relates the different indices to each other and fixes the ``energy'' parameter $c_0$. The bracket representation~\eqref{eq: bracket rep from Euler} has $3n+1$ indices and $2n+1$ bracket equations, the rank of the system is $n$, i.e.~we must leave $n$ indices free. This means, for each index $j$ we can choose either one of the parameters $\{a_j,b_j,c_j\}$ to be free. As we can see from the series above in combination with the third equation, we will be expanding in $(2p_j/S_n)^{c_j}$, so that $c_j$ must be positive. However, if we would fix $a_j \equiv m_j\in\mathbb N$, then the second equation would imply $c_j=-m_j +\dots$. Thus we should either choose $b_j\equiv m_j$ or $c_j\equiv m_j$. Let's look at the solutions for these options:
\begin{itemize}
    \item $b_j \equiv m_j \in \mathbb{N}$ leads to 
    \begin{equation}
        c_j = m_j-2\nu_j\,, \quad a_j = -m_j-\tfrac12+\nu_j\,, \quad c_0 = -\alpha +\nu_j -m_j +\dots \,.
    \end{equation}

    \item $c_j \equiv m_j \in \mathbb{N}$ leads to 
    \begin{equation}
        b_j = m_j+2\nu_j\,, \quad a_j = -m_j-\tfrac12-\nu_j\,, \quad c_0 = -\alpha -\nu_j -m_j +\dots \,.
    \end{equation}
\end{itemize}
Both solutions simply correspond to the usual positive and negative frequency modes that are required by the shadow symmetry, $\nu_j \leftrightarrow -\nu_j$ ($j=1, \ldots, n$). We have to sum over all the series with the same convergence region, which corresponds to the $2^n$ different modes where we can choose the signs of $\nu_j$. The series we obtain belongs to the class of Lauricella functions as well, but this time of type-$A$:
\begin{equation}
\label{eq: F_A series rep}
    \boxed{
    \begin{aligned}
        \I^{(n)}_{\alpha, \bm{\nu}} = \frac{(-1)^n\pi^{n/2}}{\Gamma(\tfrac12 \pm \bm\nu) S_n^{\alpha}} \sum\limits_{\bm\beta=\pm\bm\nu} &\Gamma(-2\bm\beta, \tfrac12+\bm\beta, \alpha +|\bm\beta|) \,\, \x^{\bm\beta} \\
        &\times F_A^{(n)} \left[\left.\begin{matrix}
        \alpha+|\bm\beta|;\,\, \tfrac12+\bm\beta \\ 1+2\bm\beta
    \end{matrix}\right\vert \x \right]\,.
    \end{aligned}
    }
\end{equation}
This is an alternative representation for the $n$-$K$ integral, with one additional layer of summation, defined in terms of the following dimensionless kinematic variables:
\begin{equation}
    x_j \equiv \frac{2p_j}{S_n} \,, \quad (j=1, \ldots, n) \,.
\end{equation}
The convergence domain of Lauricella $F_A$ functions is given by $\sum_{j=1}^n|x_j|<1$. This cannot be fulfilled in the physical kinematic region where by definition of the variables, we have $|x_1|+\cdots+|x_n| = 2$. Thus, the Euler representation only provides us with alternative series solutions outside the physical region. On the positive side, more is known about analytic continuation of $F_A$ compared to $F_C$, see e.g.~\cite{Srivastava1985MultipleGH}, and it would be interesting to exploit this to analytically continue our result into the physical region.

%-------------------------------------
%-------------------------------------
%-------------------------------------
\section{Analytic Continuation to the Physical Region}
\label{sec: analytic continuation}

In this section, we derive series representations of multiple-$K$ integrals that actually converge in the physical region. As we have seen, using different integral representations allows us to find various series solutions, but none provides access to all the physical kinematic configurations.

\subsection{Inductive construction}

Here, we introduce a new set of kinematic variables that allows us to construct, recursively, a series representation for multiple-$K$ integrals converging throughout the physical region.

\subsubsection{New kinematic variables}
In order to overcome the convergence issues in the physical region, we switch to a new set of dimensionless kinematic variables. For this purpose, we order the momenta
\begin{equation}
\label{eq: momentum ordering}
    0<p_n\leq \cdots \leq p_2 \leq p_1 \,,
\end{equation}
which can be achieved by permuting the indices. We then define the partial sums:
\begin{equation}
    S_j \equiv \sum\limits_{\ell =1}^j p_\ell\,,\qquad (j=2,\ldots,n)\,,
\end{equation}
and introduce the variables 
\begin{equation}
    r\equiv \frac{p_1}{S_2}\,,\qquad q_j \equiv \frac{p_j}{S_{j-1}}\,,\qquad (j=3,\dots,n)\,.
\end{equation}
Of course we can recover the original coordinates via the inverse transformations
\begin{equation}
    p_1 = r S_2\,, \qquad p_2=(1-r)S_2\,,\qquad p_j=q_jS_{j-1}\,, \qquad S_j=(1+q_j)S_{j-1}\,.
\end{equation}
These recursive inversion formulas then imply the explicit relations
\begin{equation}
    S_j = S_2\prod_{\ell=3}^j(1+q_\ell),
  \qquad
  p_j
  =
  S_2 q_j\prod_{\ell=3}^{j-1}(1+q_\ell)\,,
\end{equation}
for $j=3,\dots, n$. Since we have ordered the momenta, the new variables fulfil the bounds $\frac12 \leq r <1$ and $0\leq q_j<1$. These conditions will be crucial to obtain a convergent series representation.

\subsubsection{Recursion relation}

The ordering of the momenta allows us to write down a recursion relation for the $n$-$K$ integral. To see this, we expand Bessel-$K$ functions in terms of Bessel-$I$ functions and use their Taylor series around the origin~\eqref{eq: K/I connection formula and I series}. We write 
\begin{equation}
\label{eq:K-I-expansion-general}
  K_{\nu}(x) = \frac{\pi}{2\sin(\pi\nu)}
  \sum_{\sigma=\pm}(-\sigma)
  \sum_{m=0}^\infty
  \frac{(x/2)^{2m+\sigma\nu}}
  {m!\,\Gamma(m+1+\sigma\nu)}\,,
\end{equation}
for generic, non-integer $\nu$. Formally, expanding the outermost Bessel function $K_{\nu_n}(p_nz)$ in our integral allows us to write the following recursion relation:
\begin{equation}
\label{eq:nK-inductive-step}
    \begin{aligned}
        \I_{\alpha, \bm{\nu}}^{(n)}
        (\p) &= \frac{\pi}{2\sin(\pi\nu_n)}
        \sum_{\sigma_n=\pm}(-\sigma_n)
        \sum_{m_n=0}^\infty
        \frac{(p_n/2)^{d_n}}{m_n!\,\Gamma(m_n+1+\sigma_n\nu_n)} \\
        &\times \mathcal I_{\alpha+d_n, \{\nu_1,\ldots,\nu_{n-1}\}}^{(n-1)} (p_1,\ldots,p_{n-1})\,,
    \end{aligned}
\end{equation}
where $d_n \equiv 2m_n +\sigma_n\nu_n$. Let us inspect convergence of the recursion relation~\eqref{eq:nK-inductive-step}. At large $z\to \infty$, we have
\begin{equation}
    \prod_{j=1}^{n-1} K_{\nu_j}(p_j z) \sim z^{-(n-1)/2} e^{-S_{n-1}z} \,,
\end{equation}
whereas either resummed $I_{\pm \nu_n}(p_n z)$ branch grows as $e^{p_n z}$ up to powers of $z$. Term-wise integration is therefore valid when $p_n<S_{n-1}$, i.e.~$q_n<1$.

\vskip 4pt
In principle, one could use this recursion to push through to the first Bessel function and then formally evaluate the time integral using the method of brackets. This is equivalent to the strategy followed in~\ref{subsubsec: from schlafi series} which leads to the usual series representation in terms of Lauricella $F_C$.  As we will see, the winning strategy is to {\it not} expand two of the Bessel functions. 

\subsubsection{Two-$K$ integral \& convergent series}

To initialise the recursive construction, we consider the dimensionless terminal two-$K$ function:
\begin{equation}
    \G_{\lambda, \{\nu_1,\nu_2\}} (r) \equiv \int_0^\infty \d z\, z^{\lambda-1} K_{\nu_1}(rz) K_{\nu_2}((1-r)z)\,.
\end{equation}
As we show in the insert below, this terminal integral can be evaluated in closed form using the method of brackets:
\begin{equation}
\label{eq: terminal two-K integral}
    \begin{aligned}
        \G_{\lambda, \{\nu_1,\nu_2\}} (r) &= \frac{2^{\lambda-3}}{\Gamma(\lambda)} r^{-\lambda-\nu_2}(1-r)^{\nu_2} \Gamma\left[\frac{\lambda\pm\nu_1\pm\nu_2}{2}\right] \\
        &\times {}_2F_1\left[\left.\begin{matrix} \frac{\lambda+\nu_1+\nu_2}{2},\,  \frac{\lambda-\nu_1+\nu_2}{2}\\ \lambda \end{matrix}\right\vert 1- \frac{(1-r)^2}{r^2}\right] \,.
    \end{aligned}
\end{equation}
For $r<1/2$, it is numerically preferable to use the manifest symmetry
\begin{equation}
    \G_{\lambda, \{\nu_1,\nu_2\}} (r) = \G_{\lambda, \{\nu_2,\nu_1\}} (1-r) \,,
\end{equation}
so that the argument of the defining Gauss series always lies in $[0, 1)$. This is indeed the case since $1-(1-r)^2/r^2 = 1-(p_2/p_1)^2\leq 1$ for our choice of momentum ordering~\eqref{eq: momentum ordering}. Applying the recursion relation~\eqref{eq:nK-inductive-step} successively to $K_{\nu_n}, K_{\nu_{n-1}}, \ldots, K_{\nu_3}$ leads to the following closed-form series representation for multiple-$K$ integrals:
\begin{equation}
\label{eq: multiple-K convergent series}
    \boxed{
    \begin{aligned}
        \I_{\alpha, \bm{\nu}}^{(n)} = S_2^{-\alpha} &\prod_{j=3}^n \frac{\pi}{2\sin(\pi\nu_j)} \sum_{\sigma_n=\pm} (-\sigma_n) \sum_{m_n=0}^\infty \cdots \sum_{\sigma_3=\pm} (-\sigma_3) \sum_{m_3=0}^\infty \\
        &\times \left[\prod_{j=3}^n \frac{(\kappa_j/2)^{d_j}}{m_j! \Gamma(m_j+1+\sigma_j\nu_j)}\right] \, \G_{\lambda, \{\nu_1, \nu_2\}}(r) \,,
    \end{aligned}
    }
\end{equation}
where 
\begin{equation}
    \lambda \equiv \alpha+\sum_{j=3}^n d_j \,, \quad \kappa_j \equiv q_j \prod_{\ell=3}^{j-1}(1+q_\ell) \,, \quad d_j \equiv 2m_j+\sigma_j\nu_j\,, \quad (j=3, \ldots, n)\,.
\end{equation}
At every recursive step, the condition $0<q_j<1$ ($j=3, \ldots, n$) is automatically satisfied under the ordering~\eqref{eq: momentum ordering}. For integer values of any $\nu_j$, with $j=1, \ldots, n$, the $\sigma_j=\pm$ contributions must be combined before taking the limit $\nu_j\to N_j\in \mathbb{Z}$. The apparent singularity $1/\sin(\pi\nu_j)$ then cancels. We discuss the conformally coupled limit $\nu_j=1/2$ for $j=1, \ldots, n$ in Sec.~\ref{subsubsec: conformally coupled limit}.

\begin{framed}
{\small \noindent {\it Derivation.}---In this insert, we derive~\eqref{eq: terminal two-K integral} for the terminal two-$K$ integral, generalising the computation performed in~\cite{Gonzalez:2021vqh}. We use only tools already at our disposal: the single-$K$ Mellin transform of Sec.~\ref{subsubsec: bracket solutions as analytic continuations} and the $I_{\pm\nu}$ connection formula~\eqref{eq: K/I connection formula and I series}. Repeating the bracket computation of Sec.~\ref{subsubsec: bracket solutions as analytic continuations} for a generic power $z^{s-1}$ rather than $z^0$ gives
\begin{equation}
\label{eq: single K mellin transform general s}
    \int_0^\infty \d z\, z^{s-1} K_\nu(pz) = 2^{s-2}p^{-s}\, \Gamma\left[\frac{s\pm\nu}{2}\right]\,, \quad \Re(s)>|\Re(\nu)|\,.
\end{equation}
We now expand only the first Bessel factor in $\G_\lambda$ using~\eqref{eq: K/I connection formula and I series}, and integrate term by term against
the second using~\eqref{eq: single K mellin transform general s} with $s=\lambda+2n+\sigma\nu_1$ and $p=1-r$:
\begin{equation}
    \begin{aligned}
    \G_{\lambda,\{\nu_1,\nu_2\}}(r) = \frac{\pi}{2\sin\pi\nu_1}\sum_{\sigma=\pm}(-\sigma) &\sum_{n=0}^\infty \frac{(r/2)^{2n+\sigma\nu_1}}{n!\,\Gamma(n+1+\sigma\nu_1)}\\
    &\times\int_0^\infty \d z\, z^{\lambda+2n+\sigma\nu_1-1}K_{\nu_2}\big((1-r)z\big)\,.
    \end{aligned}
\end{equation}
The $n$-sum resums immediately into a Gauss
hypergeometric function,
\begin{equation}
\label{eq: G two branch}
    \begin{aligned}
        \G_{\lambda,\{\nu_1,\nu_2\}}(r) = \frac{\pi\, 2^{\lambda-3}}{\sin\pi\nu_1} &\sum_{\sigma=\pm}\frac{(-\sigma)\, r^{\sigma\nu_1}(1-r)^{-\lambda-\sigma\nu_1}}{\Gamma(1+\sigma\nu_1)}\, \Gamma\!\left[\frac{\lambda+\sigma\nu_1\pm\nu_2}{2}\right] \\
        &\times{}_2F_1\!\left[\begin{matrix} \frac{\lambda+\sigma\nu_1+\nu_2}{2},\, \frac{\lambda+\sigma\nu_1-\nu_2}{2}\\ 1+\sigma\nu_1 \end{matrix}\,\middle\vert\, \frac{r^2}{(1-r)^2}\right] \,.
    \end{aligned}
\end{equation}
The two branches $\sigma=\pm$---inherited from the shadow symmetry $\nu_1\to-\nu_1$ of the Bessel-$I$ expansion---are individually convergent only for $r<1/2$, and must be combined and analytically continued to reach the region $r\ge\tfrac12$. This is precisely what the classical connection formula for the hypergeometric equation, relating its local solutions at $z=\infty$ to those at $z=1$~\cite[Eq.~15.8.4]{NIST}, accomplishes: applied to~\eqref{eq: G two branch} with $z=r^2/(1-r)^2$, it maps the sum over $\sigma$ onto a \emph{single} hypergeometric function of the complementary argument $1-1/z=1-(1-r)^2/r^2\in[0,1)$, collapsing the two branches into~\eqref{eq: terminal two-K integral}.
}
\end{framed}

\subsection{Some properties}

The inductive construction produces a manifestly convergent series representation of the multiple-$K$ integral $\I^{(n)}_{\alpha,\bm\nu}(\p)$ throughout the physical region, expressed in terms of the ordered momenta $p_1\geq\cdots\geq p_n>0$ and the new kinematic variables $S_2=p_1+p_2$, $r=p_1/S_2$, $q_j=p_j/S_{j-1}$ ($j=3,\dots,n$). This representation does not treat all kinematic variables on the same footing and is written in a manifestly nested form. It is therefore instructive to examine some of the properties of this solution, which we turn to now.

\subsubsection{Conformal symmetry}
\label{subsubsec: conformal symmetry}

It is not obvious \emph{a priori} that the object one ends up with still solves the conformal Ward identities. We start by showing that it does by rewriting the special conformal Ward identity directly in the new
variables.

\paragraph{Primary Ward identities.} We first record how~\eqref{eq: conformal Ward identity} simplifies when acting on a function that depends on the momenta only through their magnitudes $p_j\equiv|\p_j|$, as is the case for $\p^{\bm\nu}\I^{(n)}_{\alpha,\bm\nu}(\p)$. Writing $f(\p)$ for such a function, one has $\bm{\partial}_j f = \hat{\p}_j\, f_{,j}$ and $\partial_j^2 f = f_{,jj} + \tfrac{d-1}{p_j}f_{,j}$, where $f_{,j}\equiv\partial f/\partial p_j$. A short computation, using $(\p_j\cdot\bm{\partial}_j)(\p_j^{\,a}\, g(p_j)) = \p_j^{\,a}(g+p_jg')$ for any radial function $g$ ($a$ is a spatial index here), shows that $(\p_j\cdot\bm{\partial}_j)\bm{\partial}_j f = \p_j\, f_{,jj}$, so that
\begin{equation}
    \bm{\K}_j f = \p_j\Big[f_{,jj} + \tfrac{d-1}{p_j}f_{,j}\Big] - 2\p_j f_{,jj} + 2(\Delta_j-d)\tfrac{\p_j}{p_j}f_{,j} = -\p_j\, \mathsf{K}_j f\,,
\end{equation}
where we have defined the purely radial second-order operator
\begin{equation}
\label{eq: radial CWI operator}
    \mathsf{K}_j \equiv \partial_{p_j}^2 - \frac{2\nu_j-1}{p_j}\,\partial_{p_j}\,, \quad \text{with} \quad 2\Delta_j-d-1=2\nu_j-1\,.
\end{equation}
The special conformal Ward identity~\eqref{eq: conformal Ward identity} therefore becomes $\sum_j \p_j\, \mathsf{K}_j f = 0$. Since the $\p_j$ satisfy only the single linear relation $\sum_j\p_j=0$ imposed by momentum conservation, this vector equation is solved precisely when $\mathsf{K}_j f$ is independent of $j$, i.e.~when the \emph{primary Ward identities}
\begin{equation}
\label{eq: primary CWI}
    \left(\mathsf{K}_i - \mathsf{K}_j\right)\, \tilde\I^{(n)}_{\alpha,\bm\nu} = 0\,, \quad \text{with} \quad \tilde\I^{(n)}_{\alpha,\bm\nu} \equiv \p^{\bm\nu} \I^{(n)}_{\alpha,\bm\nu}(\p)\,,
\end{equation}
hold for every pair $1\le i<j\le n$. The dilatation Ward identity is automatically satisfied by the homogeneity property~\eqref{eq: homogeneity}, and imposes no further constraint.

\paragraph{A master identity.} The primary Ward identities~\eqref{eq: primary CWI} are solved order by order in $z$ by a single elementary fact: for any $Z_\nu$ solving the modified Bessel equation---i.e.~$Z_\nu\in\{K_\nu,I_\nu,I_{-\nu}\}$---one has
\begin{equation}
\label{eq: master identity}
    \mathsf{K}\big[p^\nu Z_\nu(pz)\big] \equiv \Big[\partial_p^2 - \frac{2\nu-1}{p}\,\partial_p\Big] p^\nu Z_\nu(pz) = z^2\, p^\nu Z_\nu(pz)\,.
\end{equation}
Applied to $\mathsf K_j$ acting on the bare integral $\tilde{\I}^{(n)}_{\alpha,\bm\nu}$,~\eqref{eq: master identity} immediately reproduces the primary conformal Ward identities: acting with $\mathsf K_j$ simply inserts a factor of $z^2$ under the integral for every $j$ alike, so $\mathsf K_j\tilde\I^{(n)}_{\alpha,\bm\nu} = \tilde\I^{(n)}_{\alpha+2,\bm\nu}$ is manifestly independent of $j$. What makes~\eqref{eq: master identity} useful beyond this formal statement is that it holds equally for $Z_\nu=I_{\pm\nu}$: every term generated when a Bessel-$K$ leg is expanded in the $I_{\pm\nu}$ basis individually satisfies the same eigenvalue equation. This is the structural reason the recursive tower preserves conformal symmetry, as we now show explicitly.

\paragraph{Two-$K$ terminal sector.} With $r=p_1/S_2$ and $S_2=p_1+p_2$, the two-$K$ terminal function reads $\tilde{\I}_{\lambda, \{\nu_1, \nu_2\}}^{(2)}(p_1, p_2) = p_1^{\nu_1} p_2^{\nu_2} \I_{\lambda, \{\nu_1, \nu_2\}}^{(2)}(p_1, p_2) = S_2^{\nu_{12}-\lambda} h(r)$ where $h(r) = r^{\nu_1} (1-r)^{\nu_2} \G_{\lambda, \{\nu_1, \nu_2\}}(r)$. The chain rule gives
\begin{equation}
    \partial_{p_1} = \partial_{S_2} + \frac{1-r}{S_2}\partial_r\,, \qquad \partial_{p_2} = \partial_{S_2} - \frac{r}{S_2}\partial_r\,.
\end{equation}
Stripping the common factor $S_2^{\nu_{12}-\lambda-2}$, the primary conformal Ward identity $(\mathsf K_1-\mathsf K_2)\tilde\I^{(2)}_{\lambda, \{\nu_1, \nu_2\}} = 0$ collapses to the following second-order differential equation in $r$:
\begin{equation}
\label{eq: primary CWI in r}
    \begin{aligned}
        &(1-2r)h''+2(\nu_{12}-\lambda-1) h' \\
        &- \frac{2\nu_1-1}{r}\left[(\nu_{12}-\lambda)h+(1-r)h'\right] + \frac{2\nu_2-1}{1-r} \left[(\nu_{12}-\lambda)h-rh'\right] = 0 \,.
    \end{aligned}
\end{equation}
Writing $\G_{\lambda, \{\nu_1, \nu_2\}}(r) = C r^{-\lambda-\nu_2}(1-r)^{\nu_2}F(\omega(r))$ with $C \equiv \tfrac{2^{\lambda-3}}{\Gamma(\lambda)}\Gamma[\tfrac{\lambda\pm\nu_1\pm\nu_2}{2}]$ a constant, and $F(\omega) = {}_2F_1[a, b; \lambda\,|\, \omega]$ with $a=\tfrac{\lambda+\nu_1+\nu_2}{2}$ and $b=\tfrac{\lambda-\nu_1+\nu_2}{2}$, so that 
\begin{equation}
    h(r) = C \, \rho(r) \, F(\omega(r)) \,, \quad \rho(r) \equiv r^{\nu_1-\lambda-\nu_2}(1-r)^{2\nu_2}\,, \quad \omega(r) \equiv 1-\frac{(1-r)^2}{r^2} \,.
\end{equation}
Substituting this form into~\eqref{eq: primary CWI in r} via the chain rule and dividing out by the overall factor $C\rho(r)$, the primary conformal Ward identity reorganises into
\begin{equation}
    \omega(1-\omega)F''(\omega) + [\lambda-(a+b+1)\omega] F'(\omega) -ab \, F(\omega) = 0 \,,
\end{equation}
where we have used $a+b=\lambda+\nu_2$ and $\omega=(2r-1)/r^2$ to eliminate $r$ in favour of $\omega$. This is precisely the Gauss's hypergeometric differential equation.

\paragraph{Proof by induction.} Eventually, let's prove that~\eqref{eq: multiple-K convergent series} satisfies the conformal Ward identities by induction. Assume, as induction hypothesis, that $(\mathsf K_i-\mathsf K_j)\tilde\I^{(n-1)}=0$ for all $1\le i<j\le n-1$. Since $\mathsf K_i,\mathsf K_j$ ($i,j<n$) do not act on $p_n$, they commute with the sum over $\sigma_n,m_n$, so $(\mathsf K_i-\mathsf K_j)\tilde\I^{(n)}=0$ is inherited term by term. What remains is the new identity $(\mathsf K_n-\mathsf K_1)\tilde\I^{(n)}=0$, which follows from~\eqref{eq: master identity} applied to $\mathsf K_n$ acting on $p_n^{\nu_n}I_{\sigma_n\nu_n}(p_nz)$ inside the $z$-integral: it produces the eigenvalue $z^2$, exactly as $\mathsf K_1$ or $\mathsf K_2$ would one level down. The base case $n=2$ established above completes the induction.

\subsubsection{Conformally coupled limit}
\label{subsubsec: conformally coupled limit}

As a non-trivial consistency check, we now specialise to the \emph{conformally coupled} case $\nu_1 = \cdots = \nu_n = 1/2$, where $\Delta_j=(d+1)/2$ for every leg and the Bessel function becomes elementary, $K_{1/2}(z) = \sqrt{\pi/2z}\,e^{-z}$. At this point, $\I^{(n)}_{\alpha,\bm\nu}(\p)$ collapses to a pure power of the total energy $S_n \equiv \sum_{j=1}^n p_j$. We can ask whether this simple, closed-form answer is correctly
reproduced once it is extracted from the nested sums~\eqref{eq: multiple-K convergent series}, rather than from the integral itself.

\vskip 4pt
The two-$K$ terminal function simplifies drastically for $\nu_1=\nu_2=1/2$ due to the identity~\cite[Eq.~(15.4.18)]{NIST}:
\begin{equation}
    \G_{\lambda, \{\frac12, \frac12\}}(r) = \frac{\pi \Gamma(\lambda -1)}{2\sqrt{r(1-r)}} \,,
\end{equation}
and pulling out every factor independent of the summation indices in~\eqref{eq: multiple-K convergent series} leaves
\begin{equation}
    \I^{(n)}_{\alpha,(\frac12,\dots,\frac12)} = S_2^{-\alpha}\left(\frac{\pi}{2}\right)^{n-1}\frac{1}{\sqrt{r(1-r)}}\sum_n\cdots\sum_3\#\,.
\end{equation}
The nested sum is then evaluated one layer at a time using the identity
\begin{equation}
\label{eq: elementary sum identity}
    \sum_{m=0}^\infty\left[\frac{\Gamma(\alpha+2m-\tfrac32)(x/2)^{2m-\frac12}}{m!\,\Gamma(m+\frac12)} - \frac{\Gamma(\alpha+2m-\tfrac12)(x/2)^{2m+\frac12}}{m!\,\Gamma(m+\frac32)}\right] = \frac{(1+x)^{\frac32-\alpha}}{\sqrt{\pi x/2}}\,\Gamma(\alpha-\tfrac32)\,,
\end{equation}
valid for any $\alpha$ and $0\le x\le1$: at $\nu_j=1/2$, the two half-integer branches $\sigma_j=\pm$ of the $I_{\pm1/2}$ expansion---which for generic $\nu_j$ would resum into a
genuine hypergeometric function---recombine instead into a single elementary power.

\vskip 4pt
The crucial observation is that evaluating one of the sums in~\eqref{eq: multiple-K convergent series} does not change the structure of the remaining sums because the connection $\Gamma$-factor $\Gamma(\lambda -1)$ simply gets shifted $\Gamma(\lambda -1) \to \Gamma(\tilde\lambda -1)$ where $\tilde \lambda$ is obtained by the change $2m_j+\sigma_j\frac12 \to -\frac12$. Therefore, for each summation layer we need to find the corresponding variable $x_j$ and the twist $\alpha_j$. We will work from inside out, starting with $j=3$. In this case, we have $x_3 = \kappa_3 =q_3$ and $\alpha_3 = \alpha +\sum_{\ell=4}^n d_\ell$. When going to the next summation layer $j=4$, the twist is shifted and we have $\alpha_4 = \alpha-\frac12 +\sum_{\ell=5}^n d_\ell$. Furthermore, the previous summation resulted in a term $(1+q_3)^{\frac32-\alpha_3 }\propto (1+q_3)^{-2m_4-\sigma_4\frac12}$. Hence:
\begin{equation}
    x_4 = \frac{\kappa_4}{1+q_3} = \frac{q_4(1+q_3)}{1+q_3} = q_4\,.
\end{equation}
Proceeding by induction, we find:
\begin{equation}
    \alpha_j = \alpha -\tfrac12 (j-3) +\sum_{\ell=j+1}^n d_\ell\,, \quad x_j =q_j\,.
\end{equation}
Evaluating the sums layer by layer leads to 
\begin{equation}
    \I_{\alpha,\{\frac12,\dots,\frac12\}}^{(n)} = S_2^{-\alpha} \left(\frac{\pi}{2}\right)^{\frac{n}{2}} \frac{S_2}{\sqrt{p_1p_2}} \Gamma(\alpha-\tfrac{n}{2}) \prod_{j=3}^n \frac{(\frac{S_j}{S_{j-1}})^{\frac{j}{2}-\alpha}}{\sqrt{\frac{p_j}{S_{j-1}}}}\,.
\end{equation}
Using the telescopic products
\begin{equation}
    \prod_{j=3}^n\left(\frac{S_j}{S_{j-1}}\right)^{-\alpha} = \left(\frac{S_n}{S_2}\right)^{-\alpha}\,, \quad \prod_{j=3}^n \frac{S_j^{\frac{j}{2}}}{S_{j-1}^{\frac{j-1}{2}}} = \frac{S_n^{\frac{n}{2}}}{S_2}\,,
\end{equation}
we finally obtain
\begin{equation}
    \tilde{\I}_{\alpha,\{\frac12,\ldots,\frac12\}}^{(n)} \equiv \p^{\bm{\frac12}} \, \I_{\alpha,\{\frac12,\ldots,\frac12\}}^{(n)} = \left(\frac{\pi}{2}\right)^\frac{n}{2} \frac{\Gamma(\alpha-\frac{n}{2})}{(p_1+\dots+p_n)^{\alpha-\frac{n}{2}}} \,.
\end{equation}
That an infinite tower of nested hypergeometric-type sums, organised in terms of the shape variables $r,q_3,\dots,q_n$, collapses \emph{exactly} onto a bare power of $S_n$ is a stringent, representation-independent check on the recursive construction~\eqref{eq: multiple-K convergent series}.

\subsubsection{Case study: triple-$K$ integral}
\label{subsubsec: triple-K}

Because of the special role played by the $n=3$ case---the triple-$K$ integral describes all conformal three-point functions of scalar primaries---we conclude this section by spelling out the general construction explicitly for $n=3$, where it reduces to a single layer of the recursion.The relevant kinematic variables are
\begin{equation}
\label{eq: rq-definitions}
  S_2 \equiv p_1+p_2\,, \quad r \equiv \frac{p_1}{p_1+p_2},\, \quad q \equiv \frac{p_3}{p_1+p_2} \,.
\end{equation}
Physically, $S$ fixes the overall energy scale, and $r$ and $q$ can be viewed as dimensionless ``shape'' variables: $r$ specifies how the total energy is shared
between legs $1$ and $2$, and $q$ measures the size of the third momentum relative to that of the first two combined. The inverse relations read
\begin{equation}
    p_1 = rS_2\,, \quad p_2 = (1-r)S_2 \,, \quad p_3 = q S_2 \,.
\end{equation}
In the kinematic variables $(r, q)$, the entire physical region---fixed by the triangle inequalities on $(p_1,p_2,p_3)$---is mapped onto the unit square:
\begin{equation}
    0 \leq r \leq 1 \,, \quad 0 \leq q \leq 1 \,.
\end{equation}
The recursive solution~\eqref{eq: multiple-K convergent series} for $n=3$ corresponds to an expansion around the soft limit $q\to0$. Its radius of convergence extends all
the way to the opposite edge of the physical region, $q\to1$, which corresponds to the folded configuration $p_1+p_2=p_3$ where the triangle degenerates. The triple-$K$ integral is then given by
\begin{equation}
\label{eq: triple-K convergent series}
    \boxed{
    \begin{aligned}
        \I_{\alpha,\boldsymbol\nu}^{(3)} = \frac{\pi S_2^{-\alpha}}{2\sin(\pi\nu_3)} &\Bigg[\left(\frac q2\right)^{-\nu_3} \sum_{n=0}^\infty \frac{(q^2/4)^n}{n!\,\Gamma(n+1-\nu_3)} \, \G_{\alpha+2n-\nu_3, \{\nu_1, \nu_2\}}(r) \\
        &- \left(\frac q2\right)^{\nu_3} \sum_{n=0}^\infty \frac{(q^2/4)^n} {n!\,\Gamma(n+1+\nu_3)} \, \G_{\alpha+2n+\nu_3, \{\nu_1,\nu_2\}}(r) \Bigg]\,,
    \end{aligned}
    }
\end{equation}
where the two terms are the $\sigma_3=\pm$ branches of the $I_{\pm\nu_3}$ expansion of the third leg, and every coefficient is built from the terminal two-$K$ function~\eqref{eq: terminal two-K integral} evaluated at the single shape variable $r$. Unlike the Appell $F_4$ series of Sec.~\ref{subsec: triple-K integrals: recovering Appell F4}, whose convergence region $|u|+|v|<1$ misses the physical triangle configurations entirely,~\eqref{eq: triple-K convergent series} converges throughout the full physical rectangle $0\le q<1$, making it, to our knowledge, the first genuinely convergent series representation of the triple-$K$ integral in the physical region.

\begin{figure}[h!]
    %\hspace*{-1.5cm}
    \centering
    \begin{subfigure}{.5\textwidth}
        \centering
        \includegraphics[width=1\linewidth]{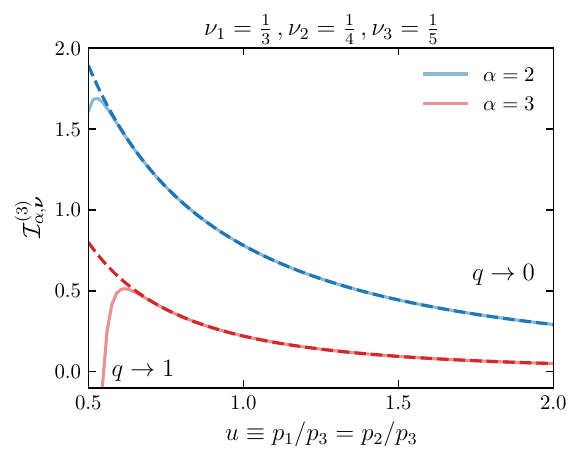}
    \end{subfigure}%
    \begin{subfigure}{.5\textwidth}
        %\vspace*{0.1cm}
        \centering
        \includegraphics[width=1\linewidth]{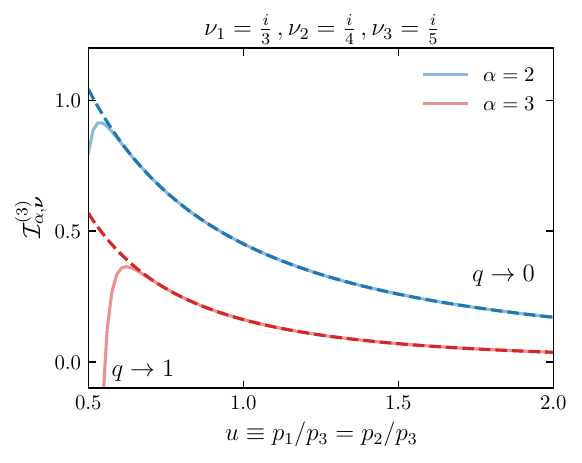}
    \end{subfigure}
   \caption{Triple-$K$ integral $\I_{\alpha, \bm{\nu}}^{(3)}$ as function of $u\equiv p_1/p_3=p_2/p_3$ with fixed $p_3$, equivalently for $0\leq q \leq 1$ at fixed $r=1/2$. {\it Left panel}: we fix the parameters $\nu_1=1/3, \nu_2=1/4, \nu_3=1/5$ (complementary series) and vary $\alpha=2, 3$. {\it Right panel}: we fix the parameters $\nu_1=i/3, \nu_2=i/4, \nu_3=i/5$ (principal series) and vary $\alpha=2, 3$. In both panels, dashed lines correspond to numerical integration and solid lines to series representations for which we have only retained $10$ terms.}
  \label{fig: tripleK}
\end{figure}

In Fig.~\ref{fig: tripleK}, we display the triple-$K$ integral $\I_{\alpha, \bm{\nu}}^{(3)}$ evaluated along the diagonal $u\equiv p_1/p_3=p_2/p_3$ in the physical region (see Fig.~\ref{fig: u/v kinematic regions}), for different values of the $\bm{\nu}$ parameters (both complementary and principal series) and the twist $\alpha$. The series representation~\eqref{eq: triple-K convergent series} yields a convergent representation throughout the entire physical domain, thereby providing a successful analytic continuation. We further observe that convergence becomes slower as one approaches $u\to1/2$, i.e.~$q\to1$, close to the boundary of the physical region. This region of degenerate (folded) triangle requires summing a larger number of terms. We have checked that the mismatch between the series representation and the exact numerical evaluation close to the boundary $q\to1$ is progressively suppressed as more terms are included in the summation. For example, one needs $\O(100)$ terms for $\alpha=3$ to reach satisfactory precision. Importantly, we have also observed that the series solution is much faster than brute-force numerical evaluation.

%-------------------------------------
%-------------------------------------
%-------------------------------------
\section{Conclusion}
\label{sec: conclusion}

Conformal symmetry is unreasonably restrictive: in position space it fixes two- and three-point functions of primary operators up to a handful of constants, and in momentum space it recasts these same objects as solutions of the corresponding conformal Ward identities, i.e.~generalised hypergeometric series, triple- and multiple-$K$ integrals. What has remained elusive is a series representation of these solutions that converges where physics actually lives, in the kinematic region defined by momentum conservation. This tension between symmetry, which hands us the rigid answers for correlators, and kinematics, which conceals it behind non-convergent representations, is what this paper set out to resolve. 

\vskip 4pt
We used the method of brackets to systematically explore the space of series representations available to multiple-$K$ conformal integrals. Applied directly, it reproduces the familiar Appell $F_4$ representation for triple-$K$ integrals, and type-$A$ and type-$C$ Lauricella representations for the general multiple-$K$ integrals, none of which converges in the physical region. By instead seeding a recursion with a two-$K$ integral and iterating the Bessel-$I$ connection formula, together with a new set of kinematic variables, we arrived at a new representation, Eq.~\eqref{eq: multiple-K convergent series}, that converges throughout the entire physical domain. This new representation satisfies the conformal Ward identities by construction, and collapses correctly onto a bare power of the total energy in the conformally coupled limit after non-trivial manipulations. Specialised to three points, it reduces to the compact, two-branch series~\eqref{eq: triple-K convergent series}.

\vskip 4pt
It would be worthwhile to revisit these integrals through the lens developed for ordinary Feynman integrals. Multiple-$K$ integrals are naturally multivariable hypergeometric functions of the Gel’fand, Kapranov \& Zelevinsky (GKZ) type~\cite{Gelfand:1990bua, Gelfand1989HypergeometricFA, Gelfand1991HYPERGEOMETRICFT, alma991007079196806161}. This formalism offers a systematic language for organising their structure, trading analytic properties for geometric ones about the associated Newton polytope---including the study of resonances, the non-generic configurations of parameters that signal reducibility to lower-dimensional subsystems. It is also tempting to ask whether canonical differential equations and Landau analysis, tools honed for controlling the singularities of Feynman integrals, apply to multiple-$K$ integrals with little modification. If so, the physical-region series constructed here by hand would emerge instead as one instance of a more systematic account of the singularities of conformal integrals and of the kinematic variables best adapted to them. 

\vskip 4pt
We close with a more personal remark. The method of brackets is older than most of the machinery now standard in the Feynman integral literature, and requires nothing beyond Ramanujan's master theorem and a modest amount of linear algebra. Yet it remains largely unknown and, in our opinion, under-used. Its virtue lies precisely in this elementariness: it renders the entire space of series representations mechanically explorable, since each choice of free index in the bracket equation is a different analytic continuation waiting to be picked out. We suspect a good number of integrals are waiting for exactly this kind of treatment.

\paragraph{Acknowledgments.} DW is funded by the Deutsche Forschungsgemeinschaft (DFG, German Research Foundation) under Germany’s Excellence Strategy---EXC-2094/2-390783311. The research of PR and DW is funded by the European Union (ERC, \raisebox{-2pt}{\includegraphics[height=0.9\baselineskip]{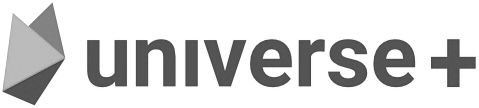}}, 101118787). {\small Views and opinions expressed are however those of the author(s) only and do not necessarily reflect those of the European Union or the European Research Council Executive Agency. Neither the European Union nor the granting authority can be held responsible for them.}

%-------------------------------------
\bibliographystyle{JHEP}
\bibliography{references.bib}

%-------------------------------------
\end{document}